\documentclass[referee]{raa}            

\usepackage{graphicx,times}             

\usepackage{graphicx,times}
\usepackage{natbib}
\usepackage{amssymb,amsmath}
\bibpunct{(}{)}{;}{a}{}{,}

\usepackage{longtable}
\usepackage{multirow}
\usepackage{booktabs}
\usepackage{pgffor}

\usepackage[a4paper=true,pagebackref=true]{hyperref}
\hypersetup{colorlinks = true, linkcolor = green, anchorcolor = red, citecolor = blue, filecolor = red, urlcolor = red}

\begin{document}

\title{Spot and Facula Activity Variations of the Eccentric Detached Eclipsing Binary KIC 8098300 Based on the Time-series Orbital Solutions
}

   \volnopage{Vol.0 (20xx) No.0, 000--000}      
   \setcounter{page}{1}          

   \author{Hubiao Niu
      \inst{1,2}
   \and Jianning Fu
      \inst{1}
   \and Jiaxin Wang
      \inst{1}
   \and Chunqian Li
      \inst{3,4}
   }

   \institute{Department of Astronomy, Beijing Normal University, No. 19, XinJieKouWai Street, Beijing 100875, China. jnfu@bnu.edu.cn\\
        \and
        Xinjiang Astronomical Observatory, Chinese Academy of Sciences, No. 150, Science 1-Street, Urumqi 830011, Xinjiang, China\\
        \and
        Key Laboratory of Optical Astronomy, National Astronomical Observatories, Chinese Academy of Sciences, Beijing 100012, China \\
        \and
        School of Astronomy and Space Science, University of Chinese Academy of Sciences, Beijing 10049, China
       }

   \date{Received~~2021 04 30; accepted~~2021~~05 15}

\abstract{The LAMOST spectra and $\it{Kepler}$ light curves are combined to investigate the detached eclipsing binary KIC 8098300, which shows the O'Connell effect caused by spot/facula modulation. The radial velocity (RV) measurements are derived by using the tomographic spectra disentangling technology. The mass ratio $q = K_1/K_2 = 0.812 \pm 0.007$, and the orbital semi-major axis $a\mathrm{\sin}i = 14.984 \pm 0.048\ R_\odot$ are obtained by fitting the RV curves. We optimize the binary model concerning the spot/facula activity with the code PHOEBE and obtain precise parameters of the orbit including the eccentricity $e=0.0217 \pm 0.0008$, the inclination $i=87.71\pm 0.04^\circ$, and the angle of periastron $\omega=284.1\pm 0.5^\circ$. The masses and radii of the primary and secondary star are determined as $M_1=1.3467 \pm 0.0001\ M_\odot$, $R_1=1.569 \pm 0.003\ R_\odot$, and $M_2=1.0940 \pm 0.0001\ M_\odot$, $R_2=1.078 \pm 0.002\ R_\odot$, respectively. The ratio of temperatures of the two component stars is $r_{teff}=0.924 \pm 0.001$. We also obtain the periastron precession speed of $0.000024\pm 0.000001\ \mathrm{d}\ cycle^{-1}$. The residuals of out-of-eclipse are analyzed using the Auto-Correlation Function (ACF) and the Discrete Fourier Transform (DFT). The spot/facula activity is relatively weaker, but the lifetime is longer than that of most single main sequence stars in the same temperature range. The average rotation period of the spots $P_{rot}=4.32\ d$ is slightly longer than the orbital period, which may be caused by either the migration of spots/faculae along the longitude or the latitudinal differential rotation. The activity may be spot-dominated for the secondary star and facula-dominated for the primary star. The method of this work can be used to analyze more eclipsing binaries with the O'Connell effect in the $Kepler$ field to get the precise parameters and investigate the difference of spot behavior between binaries and single stars. It's helpful for a deeper understanding of the stellar magnetic activity and dynamo theory.
}

   \authorrunning{Niu Hubiao, Fu Jianning, Wang Jiaxin, Chunqian Li }            
   \titlerunning{Spot and Facula Activity Variations of KIC 8098300}  

   \maketitle

%
%
\section{Introduction}           
\label{sec:intro}
It has been well known that the Sun has an 11 year magnetic activity cycle \citep{2010LRSP....7....1H}. During the activity cycle, the flux increases gradually and then falls back. The number and sizes of sunspots also increase and then decrease. In contrast to expectations, the intensity brighten for about 0.1\% at sunspot maximum activity, which is due to the contribution of faculae \citep{1979ApJ...234..707F, 2006Natur.443..161F}. The distribution of sunspots migrates gradually from mid-latitudes toward the equator, which produces the well-known butterfly diagram \citep{1904MNRAS..64..747M}. The Doppler imaging technique can be used to derive the approximate spot sizes \citep{1995MNRAS.275..534C, 2002AN....323..333B}. Through Doppler imaging, many very active, fast-rotating stars are found to have large, dark polar spots \citep{1983PASP...95..565V, 2009A&ARv..17..251S}, which can produce quasi-sinusoidal modulation. This modulation can be used to study the rotations \citep{2014ApJS..211...24M, 2016MNRAS.461..497B, 2017AJ....154..250L}, the lifetime of activity region \citep{2017MNRAS.472.1618G}, and magnetic activity cycles \citep{2017A&A...603A..52R, 2019A&A...622A..85N}. \cite{2017ApJ...851..116M} found a transition from spot-dominated to facula-dominated flux variation between rotation periods of 15 and 25 days, which implied that the transition was complete for stars at the age of the Sun.

Most of current studies of spot/facula activity are concentrated in single stars. Thanks to the large scale photometry surveys like $\it{Kepler}$ \citep{2010Sci...327..977B}, TESS \citep{2015JATIS...1a4003R} and $\it{Gaia}$ \citep{2001A&A...369..339P}, more and more eclipsing binaries are discovered \citep{2014bsee.confP..23M}. \cite{2017AJ....154..250L} reported rotation periods for 816 EBs with spot modulation. \cite{2003A&A...405..763G} analyzed high-resolution spectra of II Peg with Doppler imaging. They found that the spots were in high-latitude and the sizes of active regions changed significantly. The active longitude switches from one hemisphere to the other. \cite{2020MNRAS.492.3647X} presented the first Doppler images of the active K-type component of the prototypical binary star RS Canum Venaticorum. The maps showed that the spots located below a latitude of about $70^\circ$, but distributed widely in longitude. They also estimated the surface differential rotation rate of the K star. \cite{2018MNRAS.474.5534O} reconstructed maps of the K1-type subgiant component of the eclipsing binary KIC 11560447 with light-curve inversions (LCI) using $\it{Kepler}$ light curves. Spots on the K1 star showed two preferred longitudes, the differential rotation, and a 0.5-1.3-yr cycle. \cite{2020CoSka..50..481B} found longitudinal spot migration on the eclipsing binary KIC 9821078 with LCI. Spot activity behavior in binaries can also be investigated by model fitting \citep{2019ApJ...877...75P, 2019A&A...623A.107C}. \cite{2019A&A...623A.107C} modeled light curves of the eclipsing binary CoRoT 105895502, and found one short-lived ($\sim$ 40 days) but quasi-stationary spot, and one spot drifting at a rate of $2.3^\circ\ d^{-1}$. \cite{2019ApJ...877...75P} modeled light curves of the RS CVn–type eclipsing binary DV Psc. They found two active regions at longitude belts of about $90^\circ$ and $270^\circ$, with the activity cycles of 3.60 $\pm$ 0.03 yr and 3.42 $\pm$ 0.02 yr, respectively. They also found a correlation between spots and flares.

The LAMOST-$\it{Kepler}$ project \citep{2015ApJS..220...19D, 2018ApJS..238...30Z, 2020RAA....20..167F, 2020ApJS..251...15Z} performs spectroscopic follow-up observations for targets in the field of the $\it{Kepler}$ mission with low resolution and time-domain medium resolution spectra, respectively. The stellar atmospheric parameters and radial velocities (RVs) are obtained. Due to the unprecedented precision and almost continues of four years of the $\it{Kepler}$ light curves, and the sufficient phase coverage of LAMOST spectra, there is a good opportunity to precisely model the binaries, and further study the spot/facula activity on eclipsing binaries. \cite{2021MNRAS.504.4302W} and \cite{2020ApJ...905...67P} have investigated the spot activity of the short-period eclipsing binaries KIC 8301013 and KIC 5359678 by combining LAMOST spectra and $\it{Kepler}$ light curves, respectively. KIC 8098300 is a detached eclipsing binary located in the $\it{Kepler}$ field, and has been observed for 4 years by $\it{Kepler}$. The magnitude is $K_p$=12.817 mag, and the orbital period is $P=4.3059177$ d. It has also been observed by the LAMOST-$\it{Kepler}$ project from 2015 September to 2020 September. The physical parameters of this binary can be obtained from model solution by combining the light curves and spectra data. The out-of-eclipse residuals and spot/facula models can also be analyzed to understand the characteristic behavior of the spot/facula activity. 

Photometric and spectroscopic observations of KIC 8098300 are introduced in Section \ref{sec:obs}. O-C fitting and spectra disentangling are carried out in Section \ref{sec:dataanalysis}. The binary model solution is optimized for each cycle of light curves in Section \ref{sec:phoebe}. The out-of-eclipse residuals are analyzed to study the characters of active regions in Section \ref{sec:res}. The discussion and summary are given in Section \ref{sec:discussion} and Section \ref{sec:summary} respectively.

\section{Observations}
\label{sec:obs}
\begin{figure}[htb!]
\centering
\includegraphics[width=\textwidth,height=0.25\textheight]{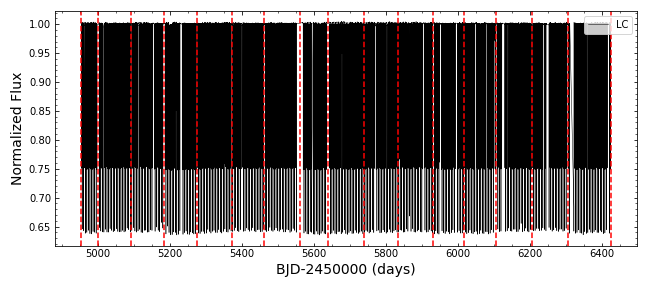}
\includegraphics[width=\textwidth,height=0.25\textheight]{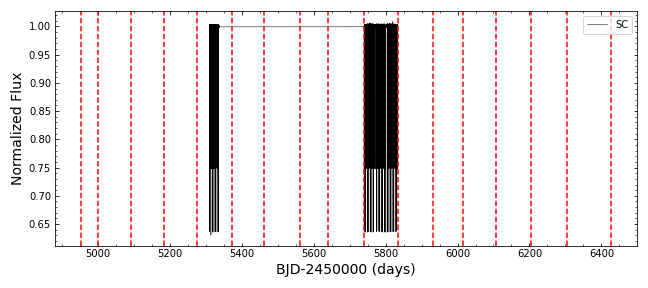}
\caption[light curves]{Light curves of KIC 8098300 from $\it{Kepler}$ photometry. The upper panel represents the long cadence data. The lower panel shows the short cadence data. The vertical red dashed lines in each panel represent the bounds of different quarters except for the first and last segments, each of which consists of two quarters data sets, i.e. Q0-Q1 and Q16-Q17, respectively.}
\label{fig:lckepler}
\end{figure}

$\it{Kepler}$ had carried out almost continuous photometric observations for 4 years (from 2009 to 2013) and obtained high-precision light curves of KIC 8098300. There are a total of 18 quarters (Q0 – Q17) of long cadence data with the sampling of 29.4 min, and 4 segments of short cadence data with the sampling of 58.8 s \citep{2011AJ....141...83P, 2011AJ....142..160S, 2016AJ....151...68K}. Those data can be accessed via the Mikulksi Archive for Space Telescopes (MAST) archive\footnote{\url{https://archive.stsci.edu/kepler/kepler_fov/search.php}}. The MAST website provides the Simple Aperture Photometry (SAP) light curves obtained from the $\it{Kepler}$ data processing pipeline and the Presearch Data Conditioning SAP (PDCSAP) light curves after detrending \citep{2012PASP..124..985S}. Based on the SAP data, the Villinova website provides the Kepler Eclipsing Binary Catalog (KEBC) and the light curves of detrending and normalization\footnote{\url{http://keplerebs.villanova.edu}} \citep{2011AJ....141...83P, 2011AJ....142..160S, 2016AJ....151...68K}. We use all of the 18 long cadence and 4 short cadence light curves accessed via the Villinova website (Fig. \ref{fig:lckepler}). Outliers are removed by sigma clipping after data smoothing and calculating the rolling mean and the standard error.

The spectra of KIC 8098300 were obtained by the LAMOST from 2015 September to 2020 September at Xinglong Station of the National Astronomical Observatories, Chinese Academy
of Sciences (NAOC). In the released LAMOST DR8 and DR9 V0-Q1 catalogs, a total of 2 low resolution spectra and 98 medium resolution spectra (49 each at the red and blue arm of the spectrometers) were obtained, respectively. The wavelength coverage of the low resolution (R $\sim$ 1800) spectrum is 370-900 nm \citep{2012RAA....12..723Z, Cui2012, 2015ApJS..220...19D}. The wavelength coverage of the medium resolution (R $\sim$ 7500) spectrum is 495-535 nm at the blue arm, and 630-680 nm at the red arm, respectively \citep{2018SPIE10702E..1IH, 2020arXiv200507210L}. Since there are more absorption lines at the blue arm, we only use the 49 spectra from the blue arm. The stellar atmospheric parameters are derived by LAMOST Stellar Parameter pipeline (LASP) using the ULySS algorithm and the ELODIE stellar library \citep{Luo2004, 2012RAA....12.1243L, Wu2014}. We adopt the average atmosphere parameters derived from the two low resolution spectra, $T_{eff1}=6425 \pm 24$ K, [Fe/H]$=-0.146\pm 0.025$. The spectra type is F5. The observations of the low and medium resolution spectra are listed in Tables \ref{tab:lowspec} and \ref{tab:medspec}, respectively. As reported by \cite{2020ApJS..251...15Z}, the RV precision increases with the increase of the signal to noise ratio (SNR). It can be better than $1\ km\ s^{-1}$ when the SNR is higher than 10, which is below the SNR of our spectra.

\begin{table}[htb!]
	\centering
	\caption{LAMOST Low Resolution Spectra.}
	\label{tab:lowspec}
	\resizebox{0.8\textwidth}{!}{%
		\begin{tabular}{cccccccc}
			\toprule[1.5pt]
			\toprule
			Date-Obs & BJD-2400000 & Phase & ExpTime & SNR & $T_{eff}$ & log $g$ & {[}Fe/H{]} \\
			  &  (d) &  & (s) &  & (K) & & \\
			\midrule
			2015-09-21 & 57287.00035 & 0.17680 & 1800 & 103.8 & 6442 $\pm$ 	
			24 & 4.216 $\pm$ 0.04 & -0.155 $\pm$ 0.023  \\
			2020-09-18 & 59111.01846 & 0.78409 & 3600 & 253.9 & 6407 $\pm$ 	
			12 & 	
			4.142 $\pm$ 	
			0.017 & -0.136 $\pm$ 	
			0.010  \\
			\bottomrule
		\end{tabular}%
	}
\end{table}

\begin{table}[htb!]
\centering
\caption{LAMOST medium resolution spectra. The expose time is $1200$ s for all the observations.}
\label{tab:medspec}
\resizebox{0.8\textwidth}{0.45\textheight}{%
    \begin{tabular}{cccccc}
	\toprule[1.5pt]
	Date & Time & BJD-2400000 & Phase  & SNR \\
	 (UT) & (UT) & (d) &  &  &   \\
	\midrule
	2018-05-31&18:12:00 & 58270.26043 & 0.52768 & 25.9 \\
	2018-05-31&18:35:00 & 58270.27640 & 0.53139 & 29.1 \\
	2018-05-31&18:59:00 & 58270.29307 & 0.53526 & 17.9 \\
	2018-05-31&19:22:00 & 58270.30904 & 0.53897 & 27.1 \\
	2019-05-21&18:38:00 & 58625.27809 & 0.97646 & 12.9 \\
	2019-05-21&19:01:00 & 58625.29406 & 0.98017 & 16.3 \\
	2019-05-21&19:25:00 & 58625.31073 & 0.98404 & 03.5 \\
	2019-06-09&17:31:00 & 58644.23226 & 0.37834 & 22.1 \\
	2019-06-09&17:55:00 & 58644.24892 & 0.38222 & 23.4 \\
	2019-06-09&18:18:00 & 58644.26490 & 0.38593 & 21.6 \\
	2019-06-09&18:42:00 & 58644.28156 & 0.38980 & 23.2 \\
	2019-06-11&17:59:00 & 58646.25177 & 0.84735 & 11.1 \\
	2019-06-11&18:23:00 & 58646.26843 & 0.85122 & 13.1 \\
	2019-06-11&18:55:00 & 58646.29066 & 0.85638 & 11.6 \\
	2019-06-14&17:40:00 & 58649.23866 & 0.54103 & 38.5 \\
	2019-06-14&18:26:00 & 58649.27061 & 0.54844 & 38.5 \\
	2019-06-14&18:49:00 & 58649.28658 & 0.55215 & 41.7 \\
	2019-06-14&19:12:00 & 58649.30255 & 0.55586 & 39.3 \\
	2020-05-31&17:32:00 & 59001.23266 & 0.28759 & 17.7 \\
	2020-05-31&18:12:00 & 59001.26044 & 0.29405 & 42.7 \\
	2020-05-31&18:38:00 & 59001.27850 & 0.29824 & 38.0 \\
	2020-05-31&19:01:00 & 59001.29447 & 0.30195 & 34.9 \\
	2020-05-31&19:25:00 & 59001.31114 & 0.30582 & 17.0 \\
	2020-06-02&17:23:00 & 59003.22648 & 0.75064 & 26.8 \\
	2020-06-02&17:46:00 & 59003.24246 & 0.75435 & 32.5 \\
	2020-06-02&18:09:00 & 59003.25843 & 0.75806 & 32.0 \\
	2020-06-02&18:33:00 & 59003.27510 & 0.76193 & 34.0 \\
	2020-06-02&18:56:00 & 59003.29107 & 0.76564 & 36.9 \\
	2020-06-02&19:19:00 & 59003.30704 & 0.76935 & 33.5 \\
	2020-06-03&17:49:00 & 59004.24457 & 0.98708 & 30.4 \\
	2020-06-03&18:12:00 & 59004.26055 & 0.99079 & 25.4 \\
	2020-06-03&18:35:00 & 59004.27652 & 0.99450 & 24.5 \\
	2020-06-03&18:59:00 & 59004.29319 & 0.99837 & 23.4 \\
	2020-06-03&19:22:00 & 59004.30916 & 0.00208 & 21.2 \\
	2020-06-05&19:06:00 & 59006.29812 & 0.46399 & 40.1 \\
	2020-06-10&18:14:00 & 59011.26217 & 0.61683 & 23.2 \\
	2020-06-10&18:38:00 & 59011.27884 & 0.62070 & 19.7 \\
	2020-06-10&19:01:00 & 59011.29481 & 0.62441 & 13.1 \\
	2020-06-10&19:25:00 & 59011.31148 & 0.62828 & 11.7 \\
	2020-06-14&16:53:00 & 59015.20604 & 0.53275 & 42.4 \\
	2020-06-14&17:16:00 & 59015.22202 & 0.53646 & 45.2 \\
	2020-06-14&17:40:00 & 59015.23868 & 0.54033 & 46.4 \\
	2020-06-14&18:13:00 & 59015.26160 & 0.54565 & 46.9 \\
	2020-06-14&18:37:00 & 59015.27827 & 0.54953 & 45.8 \\
	2020-06-14&19:00:00 & 59015.29424 & 0.55323 & 38.2 \\
	2020-06-15&17:56:00 & 59016.24982 & 0.77516 & 36.2 \\
	2020-06-15&18:19:00 & 59016.26580 & 0.77887 & 27.6 \\
	2020-06-15&18:42:00 & 59016.28177 & 0.78258 & 28.1 \\
	2020-06-15&19:06:00 & 59016.29844 & 0.78645 & 24.4 \\
	\bottomrule
    \end{tabular}%
}
\end{table}

\section{Data Analysis}
\label{sec:dataanalysis}

\subsection{O-C Analysis}
\label{subsec:o-c}

\begin{figure}[htb!]
\centering
\includegraphics[width=\textwidth,height=0.38\textheight]{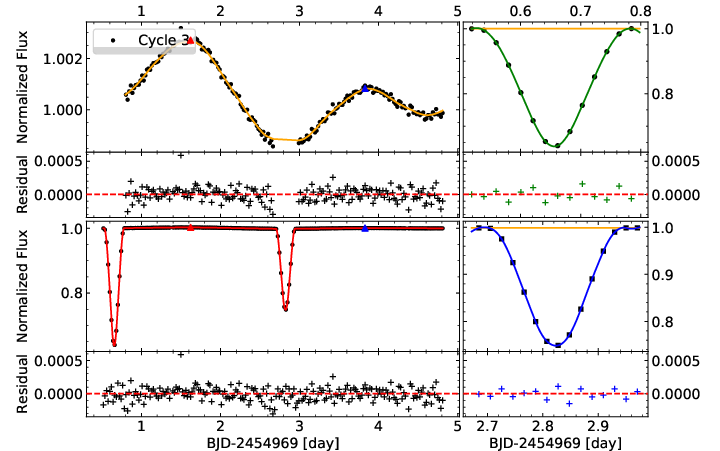}
\caption[Smoothing one cycle of the light curve]{An example of smoothing one cycle (Cycle 3) of the long cadence light curve. The upper left panel shows the smoothing and residuals of the out-of-eclipse data points, the orange curve represents the smoothing result, and the red and blue triangles indicate the positions of the primary and secondary maxima. The upper right panel shows the smoothing and residuals of the data points of the primary eclipse, while the lower right panel is the secondary eclipse. The lower left panel is the overall smoothing result and the residuals by connecting the three smoothed curves at the intersections.}
\label{fig:smooth}
\end{figure}

We segment each eclipse cycle of data into three parts, i.e. the primary eclipse, the secondary eclipse, and the out-of-eclipse. The data points of the primary and secondary eclipses are smoothed by using the $\it{LSQUnivariateSpline}$ function in the interpolate module of $\it{Python}$ package $\it{Scipy}$, respectively (Fig. \ref{fig:smooth}). Then, we take the time corresponding to the minimum of the smoothed curve as the time of minimum. The uncertainty of the time of minimum is estimated using Monte Carlo simulation by repeating the above steps 100 times, randomly adding the observation uncertainties of data points each time and finally calculating the standard deviation. We derive the reference time of the primary minimum $T_{conj}= BJD\ 2454956.737583 \pm 0.033795$ d, the orbital period $P=4.3059177 \pm 0.0000086$ d , the interval between the phase of the primary and the secondary minimum $sep=0.5034$, which are provided by the Villinova website. We fit the O-C points of the time of the primary and the secondary minima, respectively, with the linear function $T_{min}=T_0+PE$ \citep{2005ASPC..335....3S}, where $T_{min}$ represents the time of the minimum, $T_0$ is the reference time of the minimum in the ephemeris, $P$ the orbital period, and $E$ the cycle number. So the ephemeris for the time of the primary and the secondary minimum are obtained as follows:

\begin{align}
    T_{min1} &=BJD\ 2454956.73835(4) + 4.3059177(2)~E \\
    T_{min2} &=BJD\ 2454958.90556(9) + 4.3059195(5)~E
\end{align}

\begin{figure}[htb!]
\centering
\includegraphics[width=\textwidth,height=0.35\textheight]{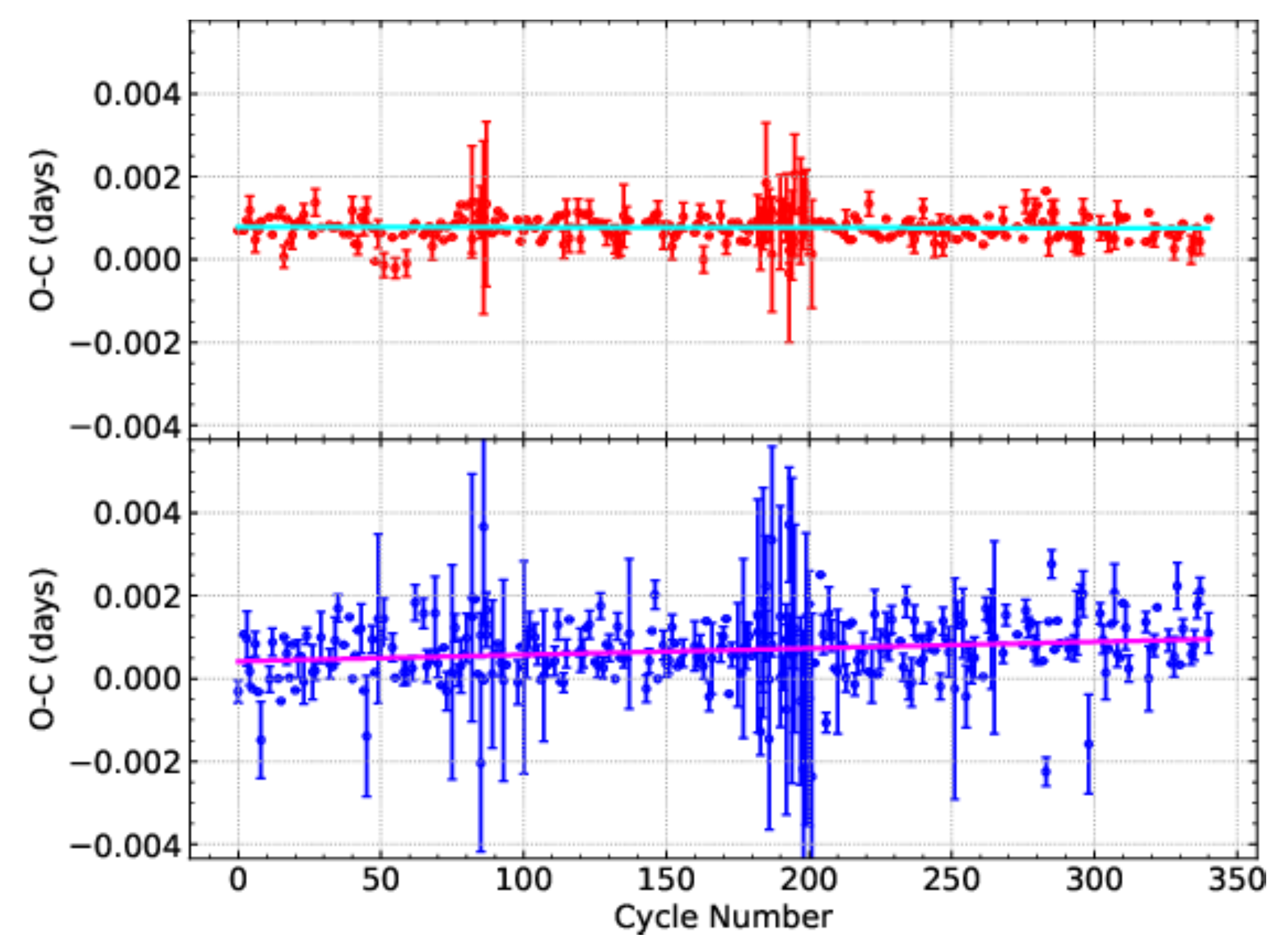}
\caption[O-C fitting]{O-C fitting. Upper panel: the red points are the time of primary minima, and the cyan solid line represents the fitting for primary minima. Lower panel: the blue points are the time of secondary minima, and the magenta solid line is the fitting of secondary minima.}
\label{fig:O-C}
\end{figure}

Compared with the ephemeris given by the Villinova website, the orbital period $P$ obtained with the O-C analysis is consistent within its uncertainty, while the difference between the reference time of the primary minimum is about and 0.00077 d (about 1.1 min) and the uncertainty is significantly reduced. Since the uncertainty of the time of primary minimum is smaller, we adopt the new ephemeris to fit the O-C points of the time of the primary minimum as the basis for subsequent analysis (Fig. \ref{fig:O-C}).

\subsection{Spectra Analysis}
\label{subsec:spectra}

The tomographic spectra disentangling technology can be applied to decompose the time series composite spectrum of binary system into the spectrum of a single sub-star, and at the same time give the orbital parameters and RVs. It bases on the principle of Doppler shift, that is, the periodic revolving of binary causes the periodic blue and red shift of the spectral lines. So no template spectrum is required \citep{1991ApJ...376..266B,1994A&A...281..286S,1995A&AS..114..393H}. We use the $\it{fd3}$ tool\footnote{\url{http://sail.zpf.fer.hr/fdbinary}} to disentangle the LAMOST medium resolution spectra. This tool is the realization of spectral tomography disentangling technology in frequency domain, which needs to provide the respective flux factors of the primary and secondary stars at different time of spectral observations \citep{2004ASPC..318..111I,2008A&A...482.1031H}. We set the flux ratio of the secondary and the primary stars $FR=F_1/F_2$ as a free parameter. Firstly, the coarse grid search is carried out with the range of 0.2-1.0 and with the step size of 0.02. At each step, the flux factors of the two component stars at each spectral observation time are calculated, respectively, under the given flux ratio $FR$ according to the phase folded and smoothed light curve, and the $\chi^2$ between the observed and the modeled spectra is also calculated after spectral disentangling. By fitting the variation of $\chi^2$ versus the flux ratio $FR$ using a fifth-order polynomial, we obtain the $FR$ corresponding to the minimum of $\chi^2$. Then, we carry out a fine grid search again with the range of $FR$ $\in$ [FR-0.1, FR+0.1] with the step size of 0.005. The uncertainty of $FR$ is estimated by the quadratic function fitting. After searching, we get the best flux ratio $FR = 0.446 \pm 0.014$ and obtain the spectrum of the primary and secondary star, respectively, with the fixed orbital period $P$ in the spectra disentangling. At the same time, we also obtain the RVs (Table \ref{tab:RVs}) and orbital parameters of the two component stars. Since the spectral disentangling technology only relies on the relative motion of the two component stars, the binary system radial velocity $\gamma$ is fixed to 0 km s$^{-1}$. We use the Monte Carlo simulation method to estimate the uncertainties of the orbital parameters and RVs by repeating the spectral disentangling process 50 times, and randomly superimposing the observation uncertainties of the observed spectra each time. Finally we calculate the standard deviation of the 50 results as the uncertainty. 

\begin{figure}[htb!]
\centering
   \includegraphics[width=\textwidth,height=0.4\textheight]{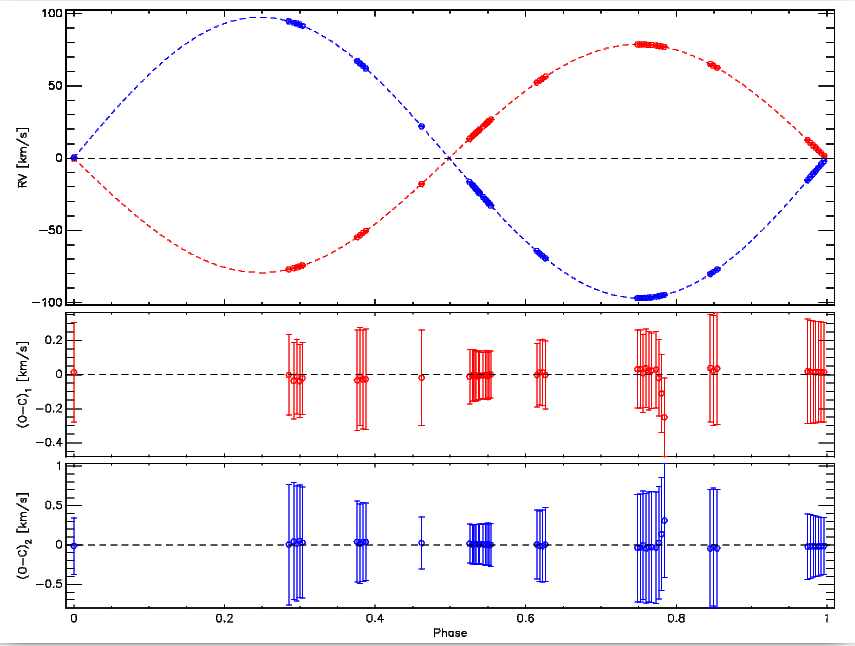}
\caption{Fitting RVs obtained from LAMOST medium spectra disentangling. The top panel shows the RVs of the primary (red points) and the secondary (blue points) star, and the orbital fitting of the primary (red dashed curve) and the secondary star (blue dashed curve). The middle and bottom panels are the corresponding residuals with uncertainties of the RVs measurements plotted.}
\label{fig:rvs}
\end{figure}

We use the IDL code $\it{rvfit}$\footnote{\url{http://www.cefca.es/people/~riglesias/rvfit.html}} \citep{2015PASP..127..567I} to fit the RVs to obtain the orbit parameters (Table \ref{fig:rvs}), i.e. the time of periastron $T_p$, the eccentricity e, the angle of periastron $\omega$, radial velocity amplitude of the primary and the secondary star $K_1 = 78.945$ km s$^{-1}$, $K_2 = 97.185$ km s$^{-1}$, the mass ratio $q = K_1/K_2 = 0.812$, and the orbital semi-major axis $a\mathrm{\sin}i = 14.984\ R_\odot$ (Table \ref{tab:binary_spot}). Since the phase coverage of the spectra is much less than that of the light curve, $K_1$ and $K_2$ obtained by spectral disentangling are relatively accurate, so the parameters $q$ and $a\mathrm{\sin}i$ are more reliable. Therefore, we fix $q$ and $a\mathrm{\sin}i$ in subsequent modeling with the light curves.

\begin{table}[htb!]
\centering
\caption{RVs Obtained from LAMOST Medium Resolution Spectra Disentangling.}
\label{tab:RVs}
\resizebox{0.7\textwidth}{0.4\textheight}{%
\begin{tabular}{ccccc}
\toprule[1.5pt]
BJD-2400000    & $RV_1$    & Uncertainty & $RV_2$     & Uncertainty \\
(d)  & (km s$^{-1})$ & (km s$^{-1})$ & (km s$^{-1})$  & (km s$^{-1})$  \\
\midrule
58270.260432 & 13.3  & 0.2 & -16.4 & 0.2 \\
58270.276404 & 15.1  & 0.2 & -18.6 & 0.2 \\
58270.293072 & 17.0  & 0.1 & -20.9 & 0.2 \\
58270.309045 & 18.7  & 0.1 & -23.1 & 0.3 \\
58625.278092 & 12.5  & 0.3 & -15.4 & 0.4 \\
58625.294065 & 10.7  & 0.3 & -13.1 & 0.4 \\
58625.310732 & 8.7   & 0.3 & -10.8 & 0.4 \\
58644.232256 & -54.7 & 0.3 & 67.3  & 0.5 \\
58644.248923 & -53.3 & 0.3 & 65.6  & 0.5 \\
58644.264896 & -51.9 & 0.3 & 63.9  & 0.5 \\
58644.281563 & -50.4 & 0.3 & 62.1  & 0.5 \\
58646.251766 & 65.1  & 0.3 & -80.2 & 0.8 \\
58646.268433 & 64.0  & 0.3 & -78.8 & 0.8 \\
58646.290656 & 62.5  & 0.3 & -76.9 & 0.7 \\
58649.238664 & 19.7  & 0.1 & -24.3 & 0.3 \\
58649.270609 & 23.2  & 0.1 & -28.6 & 0.3 \\
58649.286582 & 25.0  & 0.1 & -30.7 & 0.3 \\
58649.302555 & 26.7  & 0.1 & -32.8 & 0.3 \\
59001.232662 & -77.0 & 0.2 & 94.7  & 0.8 \\
59001.260441 & -76.1 & 0.2 & 93.7  & 0.7 \\
59001.278497 & -75.5 & 0.2 & 93.0  & 0.7 \\
59001.294470 & -75.0 & 0.2 & 92.3  & 0.7 \\
59001.311137 & -74.3 & 0.2 & 91.5  & 0.7 \\
59003.226483 & 78.7  & 0.2 & -96.9 & 0.7 \\
59003.242456 & 78.7  & 0.2 & -96.8 & 0.7 \\
59003.258429 & 78.6  & 0.2 & -96.8 & 0.7 \\
59003.275096 & 78.5  & 0.2 & -96.7 & 0.7 \\
59003.291069 & 78.4  & 0.2 & -96.5 & 0.7 \\
59003.307042 & 78.2  & 0.2 & -96.3 & 0.7 \\
59004.244575 & 7.2   & 0.3 & -8.9  & 0.4 \\
59004.260548 & 5.4   & 0.3 & -6.6  & 0.4 \\
59004.276520 & 3.5   & 0.3 & -4.3  & 0.4 \\
59004.293188 & 1.6   & 0.3 & -2.0  & 0.4 \\
59004.309160 & -0.3  & 0.3 & 0.3   & 0.4 \\
59006.298118 & -17.8 & 0.3 & 21.9  & 0.3 \\
59011.262171 & 52.2  & 0.2 & -64.3 & 0.4 \\
59011.278838 & 53.6  & 0.2 & -66.0 & 0.4 \\
59011.294811 & 54.9  & 0.2 & -67.6 & 0.5 \\
59011.311478 & 56.3  & 0.2 & -69.3 & 0.5 \\
59015.206043 & 15.8  & 0.2 & -19.4 & 0.2 \\
59015.222016 & 17.5  & 0.1 & -21.6 & 0.2 \\
59015.238683 & 19.4  & 0.1 & -23.9 & 0.3 \\
59015.261601 & 21.9  & 0.1 & -27.0 & 0.3 \\
59015.278268 & 23.7  & 0.1 & -29.2 & 0.3 \\
59015.294240 & 25.5  & 0.1 & -31.3 & 0.3 \\
59016.249824 & 77.9  & 0.2 & -95.9 & 0.7 \\
59016.265797 & 77.6  & 0.2 & -95.5 & 0.7 \\
59016.281770 & 77.3  & 0.2 & -95.1 & 0.7 \\
59016.298437 & 76.9  & 0.2 & -94.7 & 0.7 \\
\bottomrule
\end{tabular}
}
\end{table}

\section{Binary Model Solution}
\label{sec:phoebe}
%

\begin{figure}[htb!]
\centering
		\includegraphics[width=0.45\linewidth,height=0.30\linewidth]{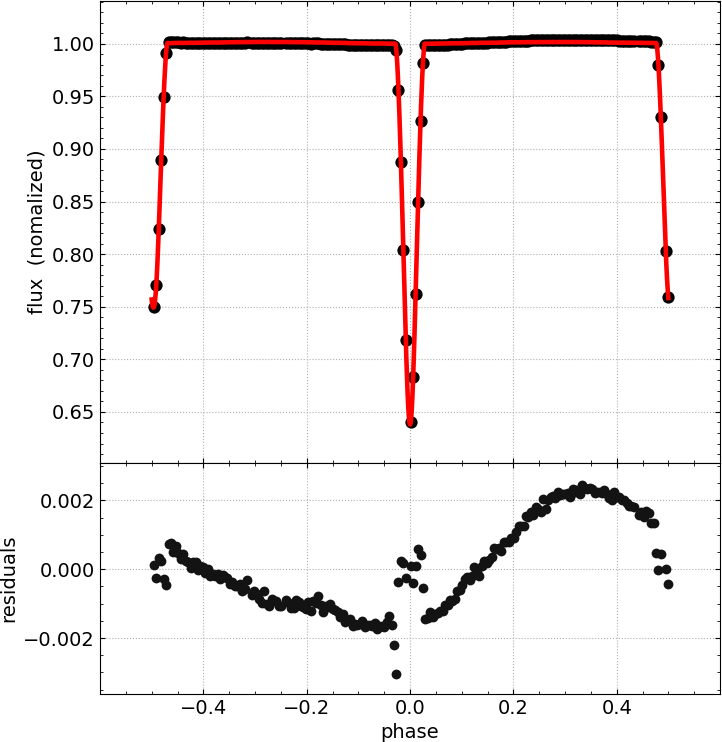}
		\includegraphics[width=0.45\linewidth,height=0.30\linewidth]{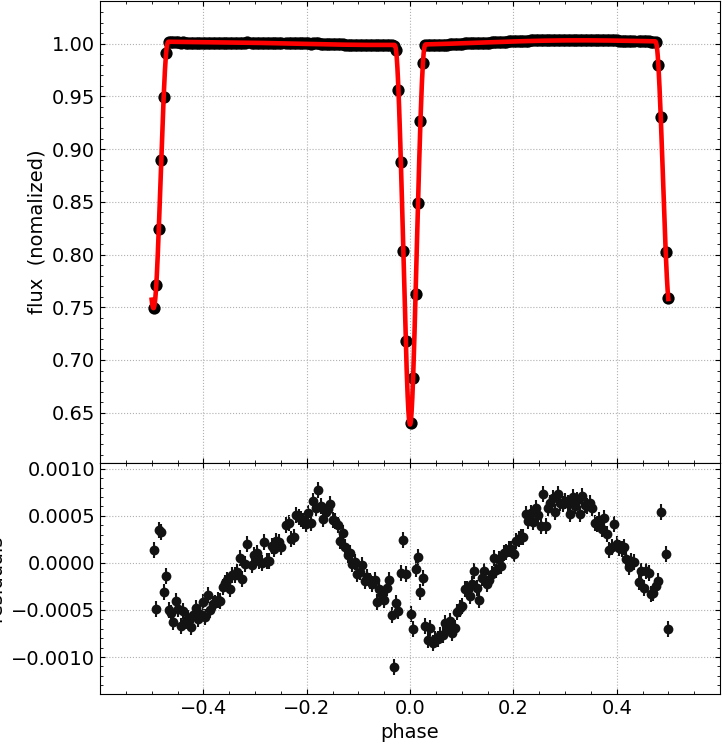}
		\includegraphics[width=0.45\linewidth,height=0.30\linewidth]{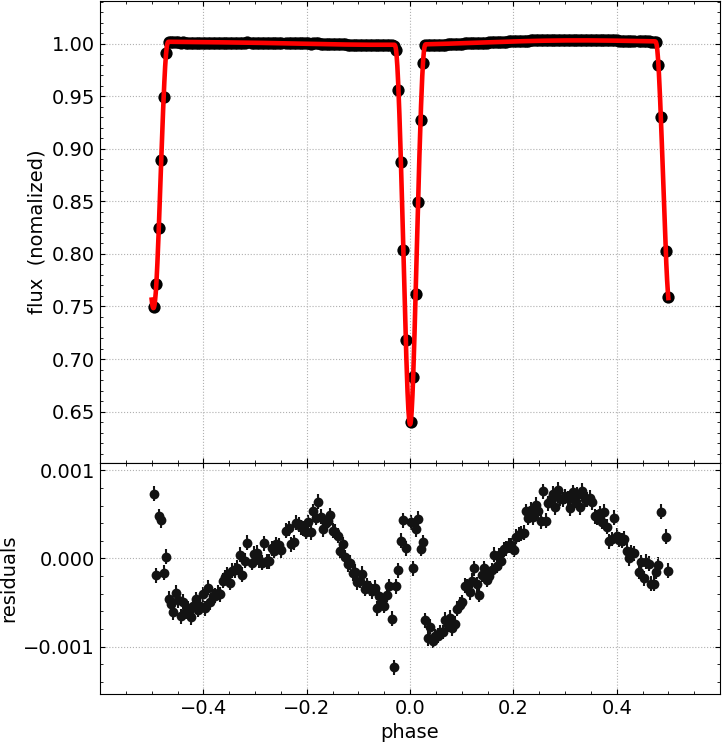}
		\includegraphics[width=0.45\linewidth,height=0.30\linewidth]{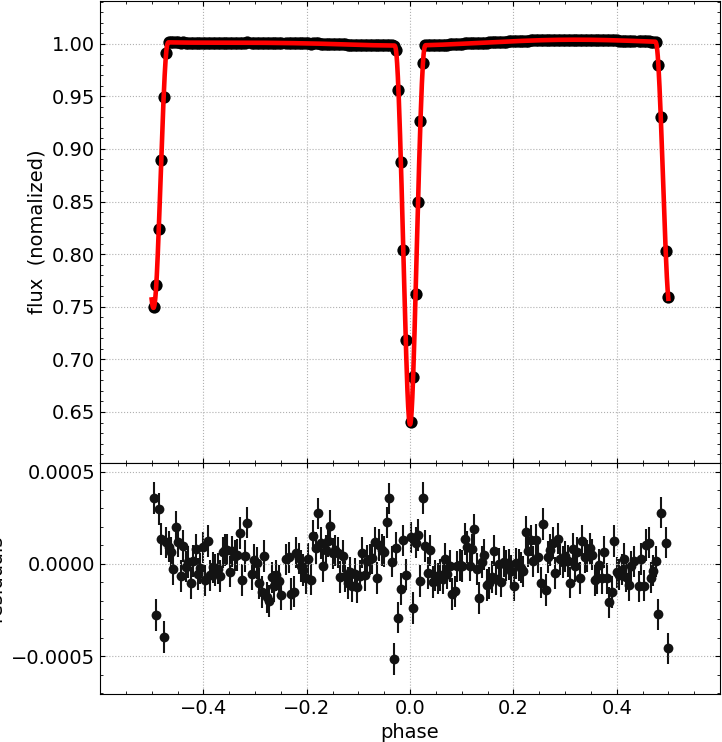}
\caption{An example of the optimal model solutions and residuals of the long cadence light curve of cycle 197. The red curve in each upper sub-panel represents the model curve produced by PHOEBE. The black points in each lower sub-panel are the corresponding residuals. Panels from left to right and top to bottom, show the model solutions without spot/facula, a spot on the primary star, a spot on the secondary star, a spot and a facula on the primary and the secondary star, respectively.}
\label{fig:lcmodel}
\end{figure}

\begin{figure}[htb!]
\centering
		\includegraphics[width=0.45\linewidth,height=0.30\linewidth]{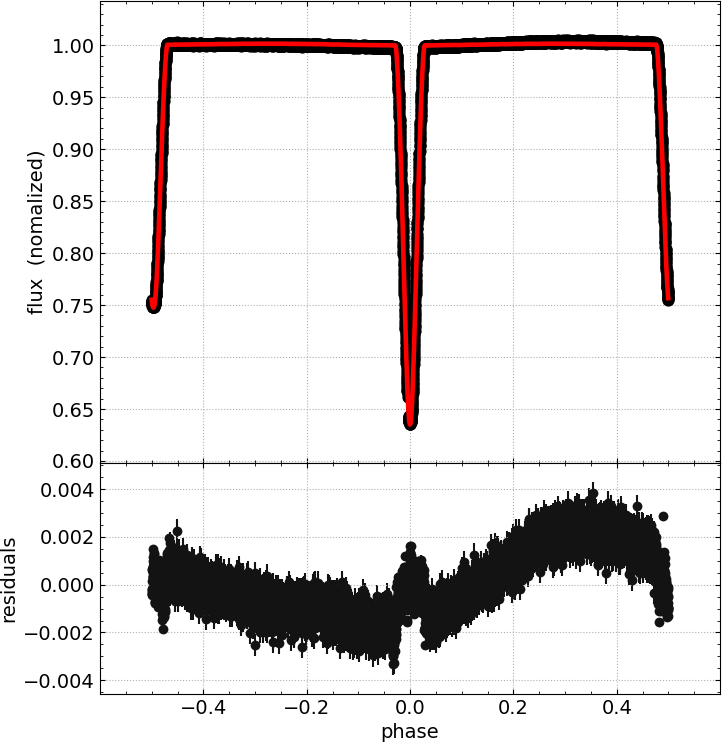}
		\includegraphics[width=0.45\linewidth,height=0.30\linewidth]{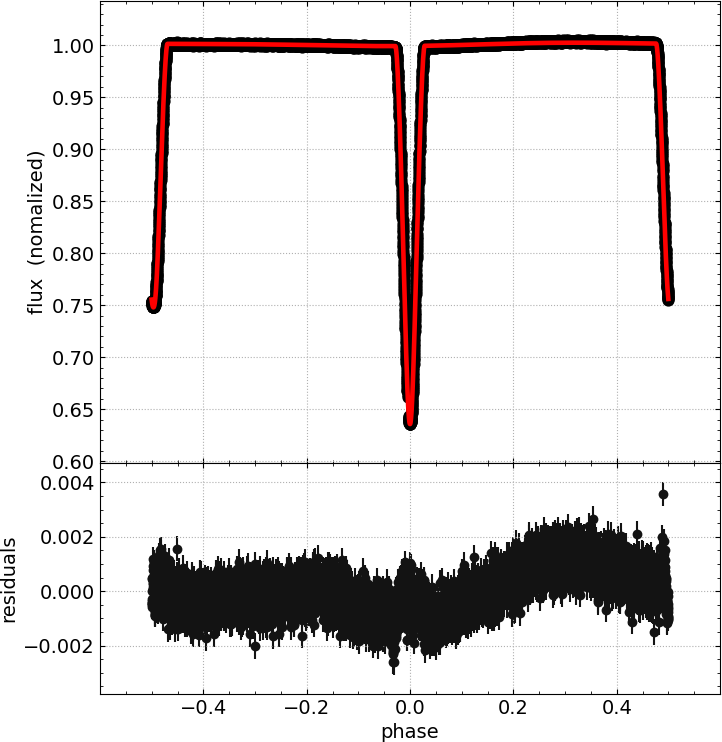}
		\includegraphics[width=0.45\linewidth,height=0.30\linewidth]{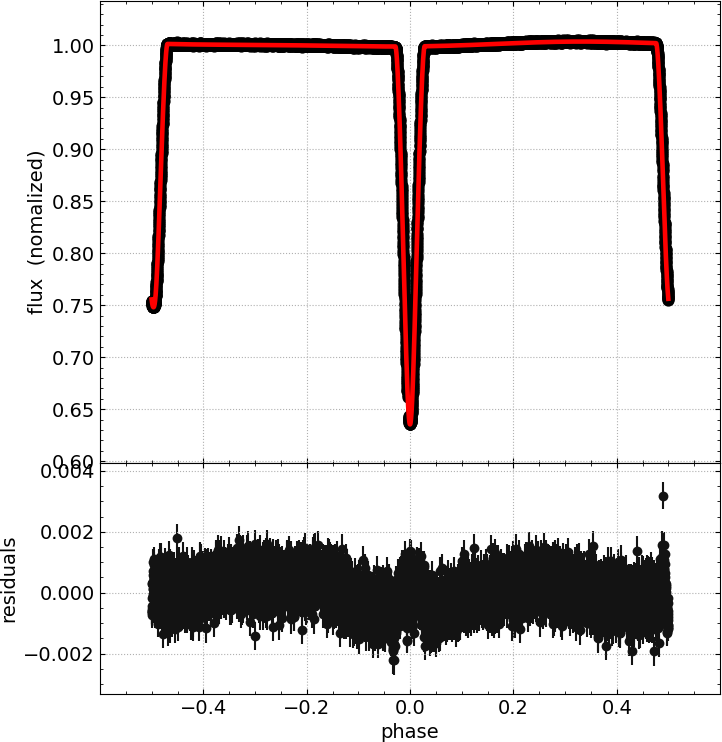}
		\includegraphics[width=0.45\linewidth,height=0.30\linewidth]{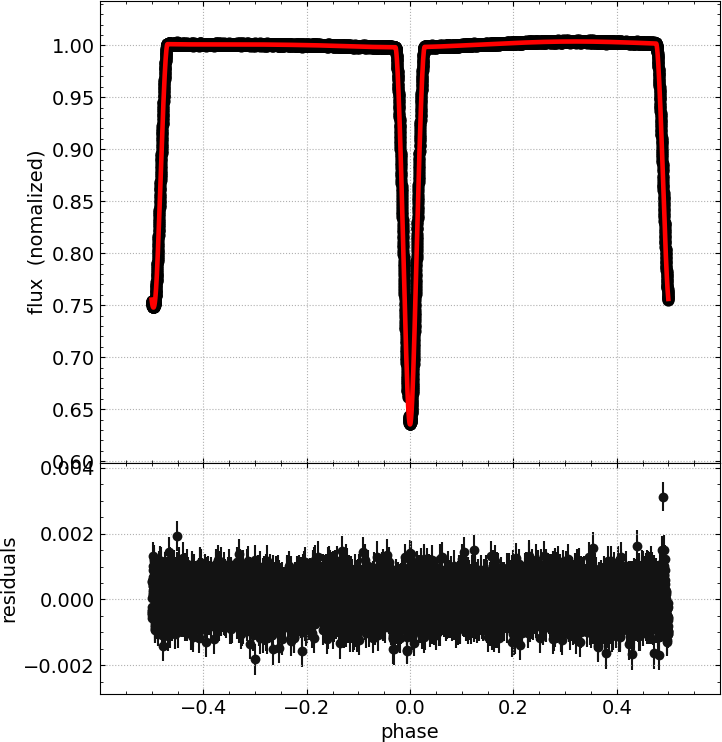}
\caption{The same as Figure \ref{fig:lcmodel} but with the short cadence data.}
\label{fig:scmodel}
\end{figure}

We use PHOEBE 2.3 (PHysics Of Eclipsing BinariEs) package \citep{2005ApJ...628..426P,2016ApJS..227...29P,2020ApJS..247...63J,2020ApJS..250...34C} to calculate the forward model for each cycle of the Kepler light curves (Figs. \ref{fig:lcmodel} and \ref{fig:scmodel}). The model is iteratively optimized using the least squares nonlinear fitting function $\it{curve\_fit}$ in the Python package $scipy.optimize$. The objective function of optimization is $\chi^2=\sum{1/\sigma^2{(f-f_{obs})}^2}$, where $f$ is the modeled flux, $f_{obs}$ is the observed flux, and $\sigma$ is the flux uncertainty. The optimal model solution is obtained corresponding to the minimum of $\chi^2$. As there are a very large number of data points in the light curves, especially for the short cadence data, running of the PHOEBE model calculation is very time-consuming. In order to reduce the calculation time, we set the grid step of the model calculation to be equivalent to the long cadence sampling interval ($\sim$ 29.4 min), and then the model fluxes corresponding to the observed points are calculated by linear interpolation. Since the duration of eclipse is very short, in order to avoid abnormal light curves caused by insufficient grid points in linear interpolation, we have increased the dense of calculation grid for the eclipse parts, with the grid step of one sixth of the long cadence sampling interval. The fixed parameters are the orbital period $P=4.3059177$ d, the time of the primary minimum $T_{conj}=\mathrm{BJD}\ 2454956.738351$ d, the mass ratio $q = 0.812$, $a\mathrm{\sin}i = 14.984\ R_\odot$, and the temperature of the primary star, $T_{eff1}=6425 \pm 24$ K, the metalicity of the two sub-stars [Fe/H] = -0.146. The value $T_{conj}$ is obtained from the O-C analysis in Section \ref{subsec:o-c}, the $q$ and $a\mathrm{\sin}i$ are obtained from the spectral disentangling  in Section \ref{subsec:spectra}, where [Fe/H], $T_{eff1}$ are calculated as the average of corresponding values given by the two low resolution spectra in LAMOST DR8 and DR9 catalogs (Table \ref{tab:lowspec}). The parameters to be optimized are $e\mathrm{\cos}\omega$, $e\mathrm{\sin}\omega$, the orbital inclination $i$, the temperature ratio of the secondary star to the primary star $r_{teff}$, the radius of the primary star $R_1$, and the radius of the secondary star $R_2$. Then, the masses of the two sub-stars can be derived. We set the initial values of the eccentricity e and the angle of periastron $\omega$ to the values obtained from the spectral disentangling in Section \ref{subsec:spectra}. In order to determine the appropriate initial values of other parameters, we use all the short cadence and long cadence data, respectively, and take the $\it{emcee}$ tool to randomly sample in a wide range of parameter space. The rough values of parameters can be guessed using the solver of PHOEBE $\it{estimator.lc\_geometry}$ and $\it{estimator.ebai}$. So the sampling range are set to a reasonable range based on the rough values and experience, $e\mathrm{\cos}\omega \in$ (-0.1, 0.1), $e\mathrm{\sin}\omega \in$ (-0.1, 0.1), $i \in$ (80.0, 90.0), $r_{teff} \in$ (0.8, 1.0), $R_1 \in$ (0.8, 2.0), $R_2 \in$(0.8, 2.0). It converges after performing 500 times of sampling by running 20 Monte Carlo Markov Chains (MCMC). Then, the average values and uncertainties of the parameters are calculated according to the posterior probability distribution to determine the initial values and the bounds of parameters to be optimized. The binary parameters are hence optimized in a narrower search range. In order to accelerate the convergence, we call the eclipsing binary light curve fast calculation code $\it{ellc}$ \citep{2016A&A...591A.111M} as the model calculation backend, which produces almost the same result as that of PHOEBE backend itself. After the optimization, we analyze the residuals and investigate the lifetime and activity of the spot/facula (see Sect. \ref{sec:res}). Finally, the binary model and spot/facula model parameters are globally optimized again in a narrower range at the same time (see Sect. \ref{subsec:spots}) for each cycle. We take the average of each parameter over all the 264 cycles as the final parameter value, and estimate the uncertainty by calculating the standard deviation of each parameter over all cycles. We compare the parameters obtained from the short cadence data and the long cadence data, and find that they are very consistent to each other within the uncertainties (Table \ref{tab:binary_spot}).

\begin{table}[htb!]
\centering
\caption{The Average Parameters Obtained from Binary and Spot/facula Model Solutions.}
\label{tab:binary_spot}
\begin{minipage}[t]{0.99\textwidth}
\resizebox{0.96\textwidth}{0.25\textheight}{%
    \begin{tabular}{c|ccc|ccc}
	\toprule
	\multirow{2}* {Parameter } & \multicolumn{3}{c|}{Short Cadence} &   \multicolumn{3}{c}{Long Cadence} \\
	\cline{2-4}  \cline{5-7}
	 & Primary & System & Secondary  & Primary & System & Secondary \\
	\midrule
	$T_{conj}$ (d) &  & $2454956.73835\pm0.00004^a$ & & & $2454956.73835\pm0.00004^a$ &   \\
	$P$ (d) &  & $4.3059177\pm0.0000002^a$ & & & $4.3059177\pm0.0000002^a$ &   \\
	$\gamma$ (km s$^{-1}$) & & $0.0^b$ & & & $0.0^b$ &  \\
	$q$   &  & $0.812\pm0.007^b$ &    &   &$0.812\pm0.007^b$  & \\
	$a\mathrm{sin}i$   &  &$10.429\pm0.048^b$  &    &   &$10.429\pm0.048^b$  & \\
	$i\ (^\circ)$ & & 87.71$\pm$0.04 & & & 87.71$\pm$0.05 &  \\
	$e\mathrm{cos}\omega$ & & 0.00529$\pm$0.00004 & & & 0.00529 $\pm$0.00006 &  \\
	$e\mathrm{sin}\omega$ & & -0.0211$\pm$0.0008 & & & -0.0209$\pm$0.0009 &  \\
	$e$ & & 0.0217$\pm$0.0008 & & & 0.0215$\pm$0.0009 &  \\
	$\omega\ (^\circ)$ & & 284.1$\pm$0.5 & & & 284.2$\pm$0.6 &  \\
	$r_{teff}$ & & 0.924$\pm$0.001 & & & 0.924$\pm$0.001 &  \\
	$R\ (R_\odot)$ &1.569$\pm$0.003& &  1.078$\pm$0.002 &1.566$\pm$0.002& & 1.078$\pm$0.004 \\
	$M\ (M_\odot)$ & 1.3467$\pm$0.0001 & & 1.0940$\pm$0.0001 & 1.3467$\pm$0.0002 &  & 1.0940$\pm$0.0001 \\
	$T_{eff}$ (K) &$6425\pm24^c$ & & 5936$\pm$25 &$6425\pm24^c$ & & 5936$\pm$25 \\
	$L_2/L_1$ &  & $0.344\pm0.009$ &  &  & $0.343\pm0.004$ &  \\
	\midrule
	$colat1\ (^\circ)$ & & &70$\pm$32& & &65$\pm$34 \\
	$long1\ (^\circ)$ & & &144$\pm$72 & & &103$\pm$61 \\
	$radius1\ (^\circ)$ & & &21$\pm$10 & & &18$\pm$11 \\
	$relteff1 $ & & &0.727$\pm$0.559 & & & 0.911$\pm$0.553 \\
	$colat2\ (^\circ)$ & 44$\pm$32& & & 47$\pm$31 & & \\
	$long2\ (^\circ)$ & 88$\pm$43 & & & 95$\pm$57 & &  \\
	$radius2\ (^\circ)$ & 34$\pm$27  & & & 30$\pm$26 & &  \\
	$relteff2 $ & 0.998$\pm$0.032 & & & 1.003$\pm$0.065 & &  \\
	\bottomrule
	\end{tabular}
} \\
\footnotesize {$^a$: Derived by O-C fitting, fixed in binary modeling. \\
$^b$: The RVs are only related to the relative motion of the two sub-stars with the method of spectral disentangling technique, so the radial velocity of the binary system is always fixed to 0 km s$^{-1}$.\\
$^c$: Obtained from LAMOST low resolution spectra.}
\end{minipage}
\end{table}

\section{Analysis of Out-of-eclipse Residuals}
\label{sec:res}

\begin{figure}[htb!]
\centering
   \includegraphics[width=\textwidth,height=0.2\textheight, angle=0]{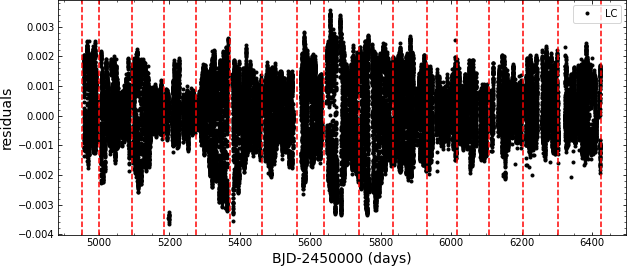}
   \includegraphics[width=\textwidth,height=0.2\textheight, angle=0]{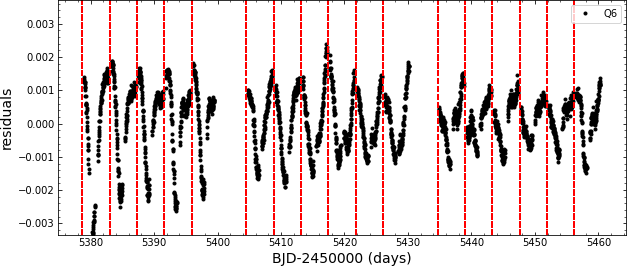}
\caption{Upper panel: the out-of-eclipse residuals (black points) for all quarters (Q0-Q17). Lower panel: the residuals zoomed for Q6. The red dashed vertical lines in the upper panel represent the bounds of different quarters except for the first and last segments, each of which consists of two quarters of data sets, i.e. Q0-Q1 and Q16-Q17, respectively. The red dashed vertical lines in the lower panel indicate the positions of the primary minima.}
\label{fig:res}
\end{figure}

Ignoring the influence of spot/facula activity, the binary model parameters obtained in the previous section, i.e. $e\mathrm{cos}\omega$, $e\mathrm{sin}\omega$, the orbital inclination $i$, the temperature ratio of the secondary star to the primary star $r_{teff}$, the radius of the primary star $R_1$, and the radius of the secondary star $R_2$, are used to calculate the theoretical fluxes and then the residuals for each cycle. It can be seen that there is an obvious quasi-sinusoidal variation in the out-of-eclipse. The primary and secondary maxima are not equal to each other (Fig. \ref{fig:res}), which is the evidence of the O'Connell effect \citep{Connell51}. This is generally considered to be modulated by the spot activity. Based on this assumption, we analyze the out-of-eclipse residuals as follows.

\subsection{Active Region Lifetime }
\label{subsec:lifetime}
\begin{figure}[htb!]
\centering
	\includegraphics[width=0.48\textwidth,height=0.21\textheight, angle=0]{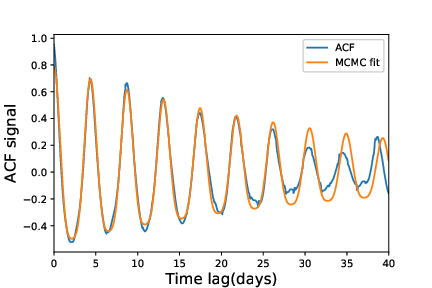}
	\includegraphics[width=0.48\textwidth,height=0.21\textheight, angle=0]{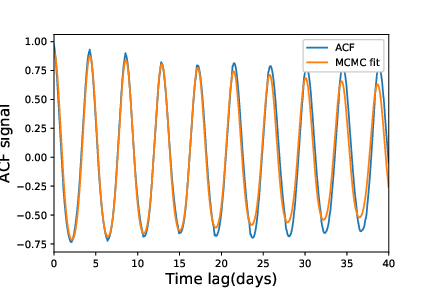}
	\includegraphics[width=0.48\textwidth,height=0.21\textheight, angle=0]{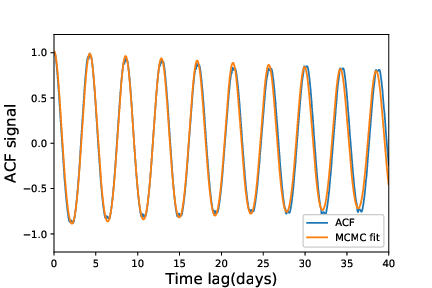}
	\includegraphics[width=0.48\textwidth,height=0.21\textheight, angle=0]{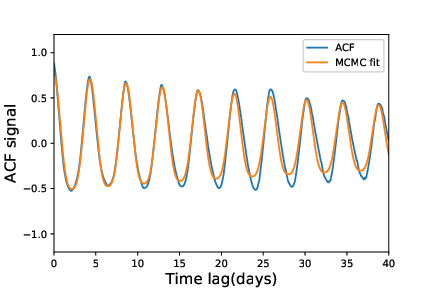}
	\caption[Active region lifetime]{Estimation of active region lifetime by fitting the ACF of out-of-eclipsing residuals using uSHO function. Panels from left to right and top to bottom, show the results from 1st to 4th year, respectively. In each panel, the blue curve is the smoothed ACF, and the orange curve represents the fitting using MCMC method.}
\label{fig:ACF}
\end{figure}

Spots keep appearing and disappearing at different latitudes and longitudes on the surface of the sub-stars, causing the modulation in the phase and amplitude of the out-of-eclipse fluxes, so the active region shows a certain lifetime. The auto-correlation function (ACF) method can be used to estimate the lifetime of the dominant active region \citep{2013MNRAS.432.1203M,2014ApJS..211...24M}. The ACF will have series of peaks at the lags close to integer multiples of the rotation period of the dominant spot, where the signal has the highest degree of self-similarity. The peak heights decrease with time delay due to spot formation and decay. The decay rate of the side lobes can be used to characterize the lifetime of the active region. The underdamped simple harmonic oscillator (uSHO) function is used to fit the ACF for each year respectively using the MCMC method \citep{2017MNRAS.472.1618G}:
\begin{equation}
y(t)=e^{(-t/\tau_{AR})}(Acos(\frac{2\pi t}{P})+Bcos(\frac{4\pi t}{P})+y_0),
\end{equation}	
where $A$ and $B$ are the amplitudes of the cosine terms and $y0$ is the offset to $y=0$. $P$ is the stellar rotation period corresponding to the time lag at which the highest peak occurs. $\tau_{AR}$ represents the decay time-scale of the ACF which indicates the lifetime of the dominant spot. The results are shown in Figure \ref{fig:ACF} and listed in Table \ref{tab:ACF}. It can be seen that the lifetime of the active region is relatively long except for the first year.

\begin{table}[htb!]
		\centering
		\caption{Active Region Lifetime}
		\label{tab:ACF}
		\resizebox{0.65\textwidth}{0.06\textheight}{%
			\begin{tabular}{cccccc}
				\toprule[1.5pt]
				\toprule
				\multirow{2}*{Year} &$\tau_{AR}$  & A & B & $y_0$ & $P_{rot}$ \\
				 & (d) & (d) & (d) & (d) & (d) \\
				\midrule
				1st & 35.4 & 0.658 & 0.138 & -0.008 & 4.364   \\
				2nd &106.8 & 0.820 & 0.099 & -0.008 & 4.295 \\
				3rd &160.7 & 0.957 & 0.064 & -0.006 & 4.277 \\
				4th &66.5 & 0.639 & 0.119 & -0.002 & 4.311 \\
				\bottomrule
			\end{tabular}%
		}
\end{table}

\subsection{Rotation Period}
\label{subsec:prot}
\begin{figure}[htb!]
\centering
\includegraphics[width=0.8\textwidth,height=0.28\textheight, angle=0]{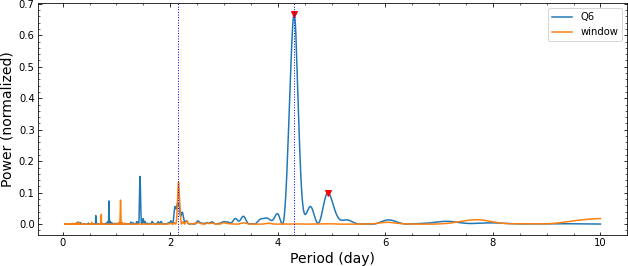}
\caption[The periodogram of Q6]{An example of the periodogram of the out-of-eclipse residuals for Q6. The blue curve is the power spectrum. The orange curve is the spectrum window. The red triangles mark the position of the significant peaks. From left to right, vertical blue dashed lines represent the half and the full orbital period, respectively.}
\label{fig:pdm_Q6}
\end{figure}

\begin{figure}[htb!]
\centering
   \includegraphics[width=0.9\textwidth,height=0.28\textheight, angle=0]{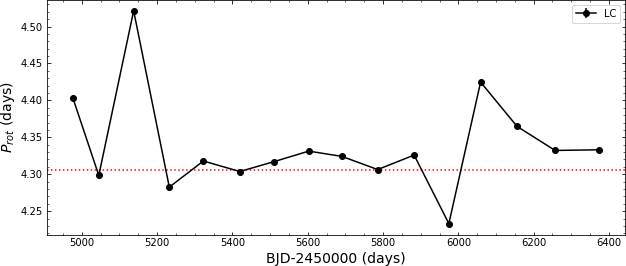}
\caption{The rotation period of spot/facula is shown for each quarter, and the horizontal dotted line represents the orbital period.}
\label{fig:prot}
\end{figure}

Due to the differential rotation, the rotation period of the spot/facula at different positions on the surface of a star would be different. For the short-period eclipsing binary, the rotation period of the sub-stars is the same as the orbital period due to the tidal synchronization. Compared with ACF, the Discrete Fourier Transform (DFT) is more conducive to analyze the rotation periods of spots/faculae. Therefore, the generalized Lomb-Scargle (LS) module in the $\it{Astropy}$ package \citep{1976Ap&SS..39..447L, 1982ApJ...263..835S, 2009A&A...496..577Z, 1989ApJ...338..277P, 2018ApJS..236...16V} is employed to analyze the residuals of long cadence light curves for each quarter (Fig. \ref{fig:pdm_Q6}).
From Figures \ref{fig:pdm_Q6} and \ref{fig:prot}, it can be seen that the rotation period of spots/faculae is close to the orbital period, with an average of $P_{rot}=4.32$ d. The result is listed in Table \ref{tab:Prot}. Most of the rotation periods are slightly longer than the orbital period (4.3059177 d), indicating that the rotation is slightly slower than that of the sub-star during most of the time, which may be caused by either the migration of spots/faculae along the longitude (Fig. \ref{fig:longitude}) or the latitudinal differential rotation.

%

\begin{table}[htb!]
\centering
\caption{The Rotation Periods of Spots/faculae for Each Quarter.}
\label{tab:Prot}
\resizebox{0.95\textwidth}{0.04\textheight}{%
\begin{tabular}{cccccccccccccccccc}
	\toprule[1.5pt]
    \toprule
	Quarter & Q0-Q1 & Q2& Q3& Q4& Q5& Q6& Q7& Q8& Q9& Q10& Q11& Q12& Q13& Q14& Q15& Q16-Q17\\
	\midrule
	$P_{rot}$ (d) & 4.403& 4.298& 4.521& 4.283& 4.318& 4.304& 4.317& 4.331& 4.324& 4.306& 4.326& 4.233& 4.425& 4.365& 4.332& 4.333 \\
	\bottomrule
\end{tabular}%
}
\end{table}

\subsection{Active Region Evolution}
\label{subsec:AR}
\begin{figure}[htb!]
\centering
\includegraphics[width=\textwidth,height=0.35\textheight, angle=0]{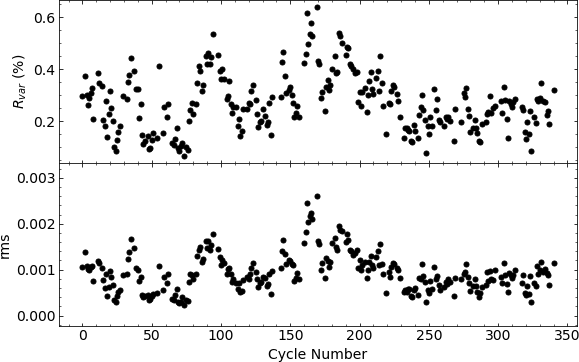}
\caption{The variability range and the root-mean-square scatter of the active region versus cycle number.}
\label{fig:rav_rms}
\end{figure}

In order to quantitatively describe the flux modulation due to active region during each eclipse cycle, two indicators $R_{var}$ and $rms$ are calculated for each cycle. $R_{var}$ represents the variability range, which is defined as the range between the 5th and 95th percentile of the sorted intensity, and $rms$ is the root-mean-square scatter \citep{2011AJ....141...20B,2017A&A...603A..52R}. Figure \ref{fig:rav_rms} shows the variations of $R_{var}$ and $rms$ versus cycle number. It can be seen that the spot activity is enhanced around the cycle 90 and 160, but the four years of observation are not long enough to obtain an explicit magnetic activity cycle.

\section{Discussion}
\label{sec:discussion}

\subsection{Periastron precession}
\label{subsec:precession}
\begin{figure}[htb!]
\centering
\includegraphics[width=\textwidth,height=0.25\textheight, angle=0]{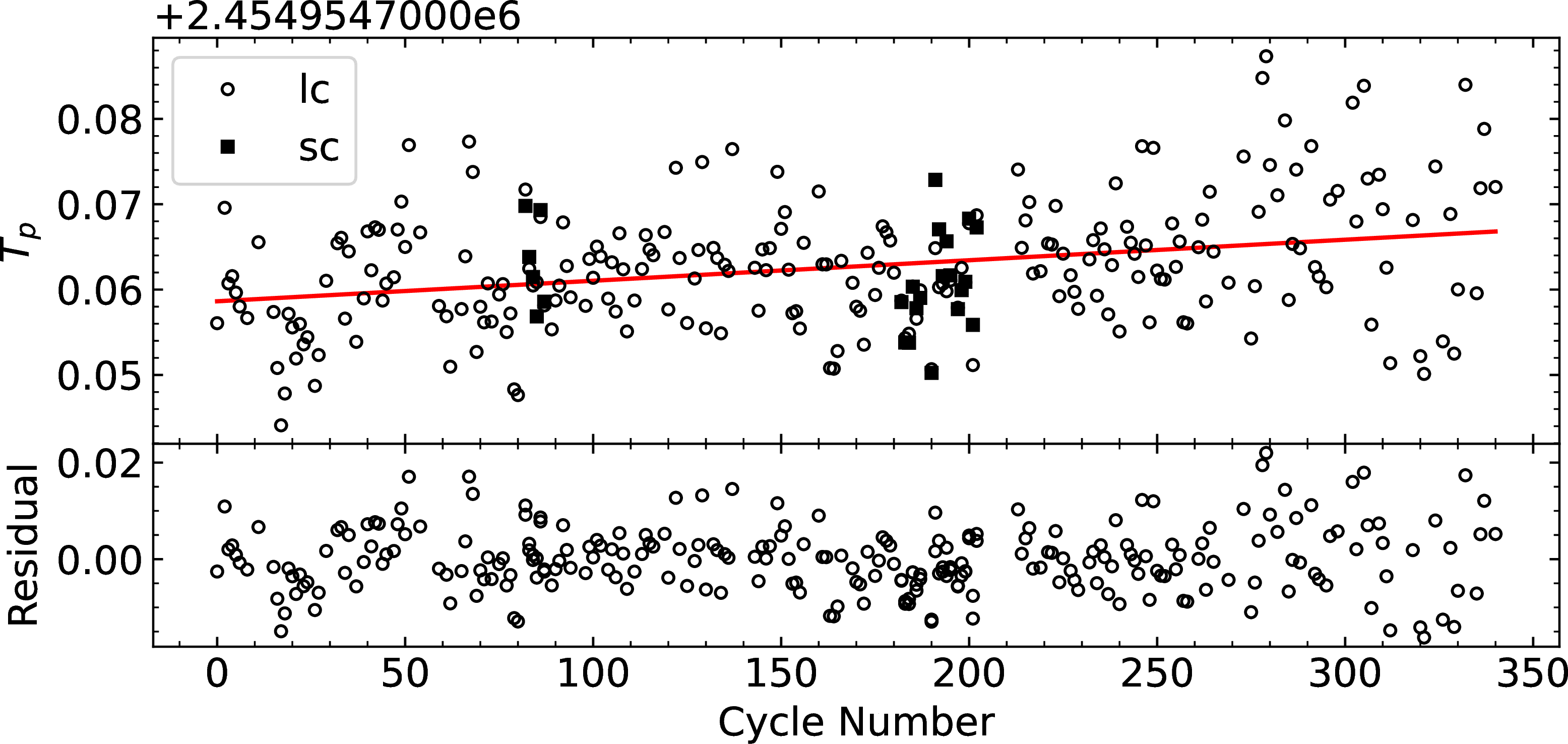}
\includegraphics[width=\textwidth,height=0.25\textheight, angle=0]{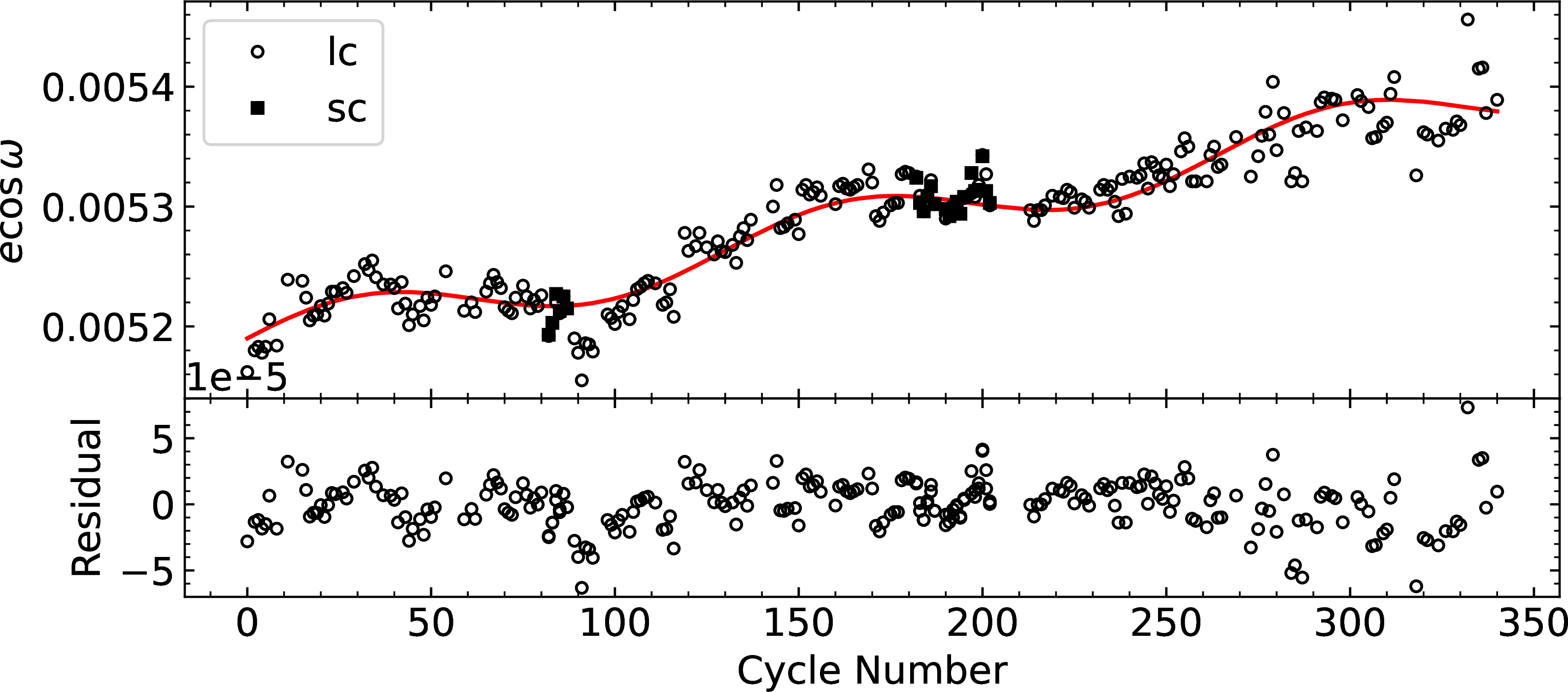}
\includegraphics[width=\textwidth,height=0.25\textheight, angle=0]{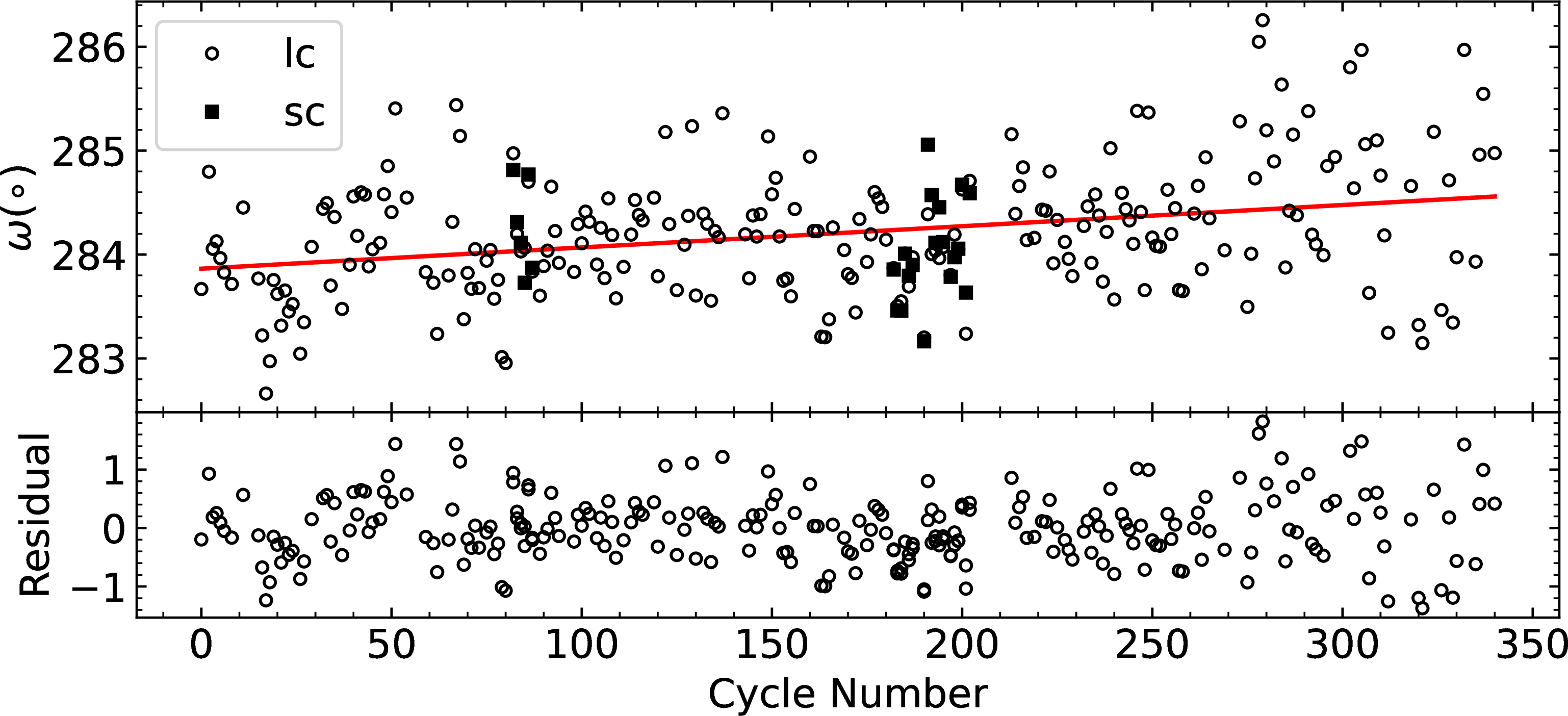}
\caption{The periastron precession. The black open circles in each upper sub-panel are the optimized binary parameters for long cadence data, while the black filled squares represent that of short cadence data. The red curve in each upper sub-panel represents the fitted curve. In the lower sub-panels, the corresponding residuals are plotted.}
\label{fig:precession}
\end{figure}

After optimization of binary model with a spot and a facula for each cycle of light curve, we finally get the parameters of KIC 8098300 that changed over time. We find the periastron procession of the binary system. As can be seen in Figure \ref{fig:precession}, the orbital parameters $T_p$, $e\mathrm{cos}\omega$ and $\omega$, increase with the cycle number. There are no trends for the rest of other parameters. This trend is the most evident for $e\mathrm{cos}\omega$. However, it also superimposes a sinusoidal signal, which may be caused by the modulation of spot/facula activity. $T_p$ and $\omega$ versus cycle number are fitted with the linear functions with the periastron  precession speeds is of $0.000024\pm 0.000001$ d cycle$^{-1}$ ($\sim$ 2.0736 min cycle$^{-1}$) or $0.0020\pm 0.0004^\circ\ cycle^{-1}$, respectively. The $e\mathrm{cos}\omega$ values are fitted with the combination of a linear and a sine function, and the periastron  precession is derived as $0.0015\pm 0.0003^\circ\ cycle^{-1}$. We hence get a modulation time scale of 134 cycles (about 578 d), which might be due to the magnetic activity cycle.

\subsection{Spot/facula model}
\label{subsec:spots}
\begin{figure}[htb!]
\centering
\includegraphics[width=\textwidth,height=0.3\textheight, angle=0]{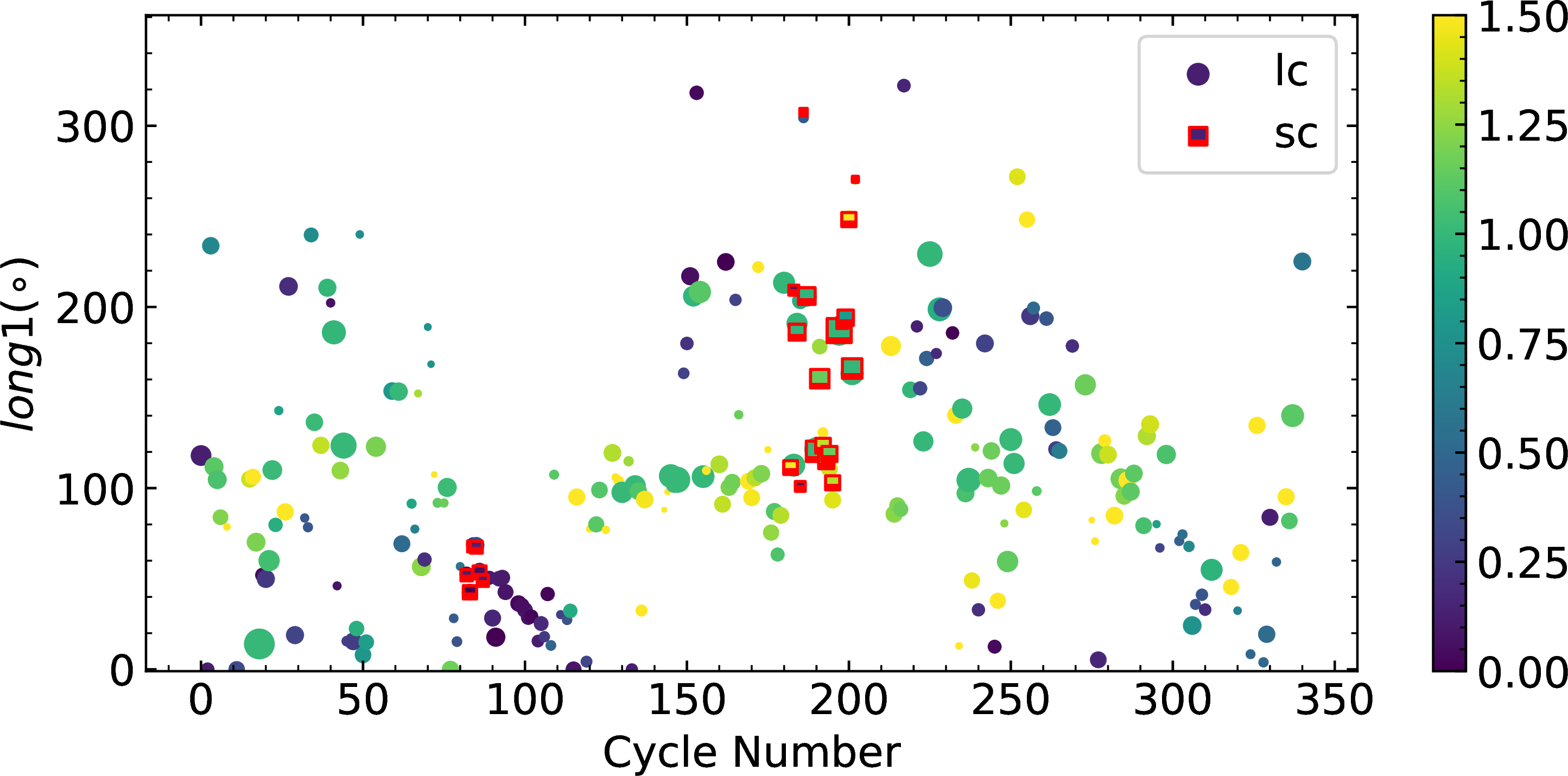}
\includegraphics[width=\textwidth,height=0.3\textheight, angle=0]{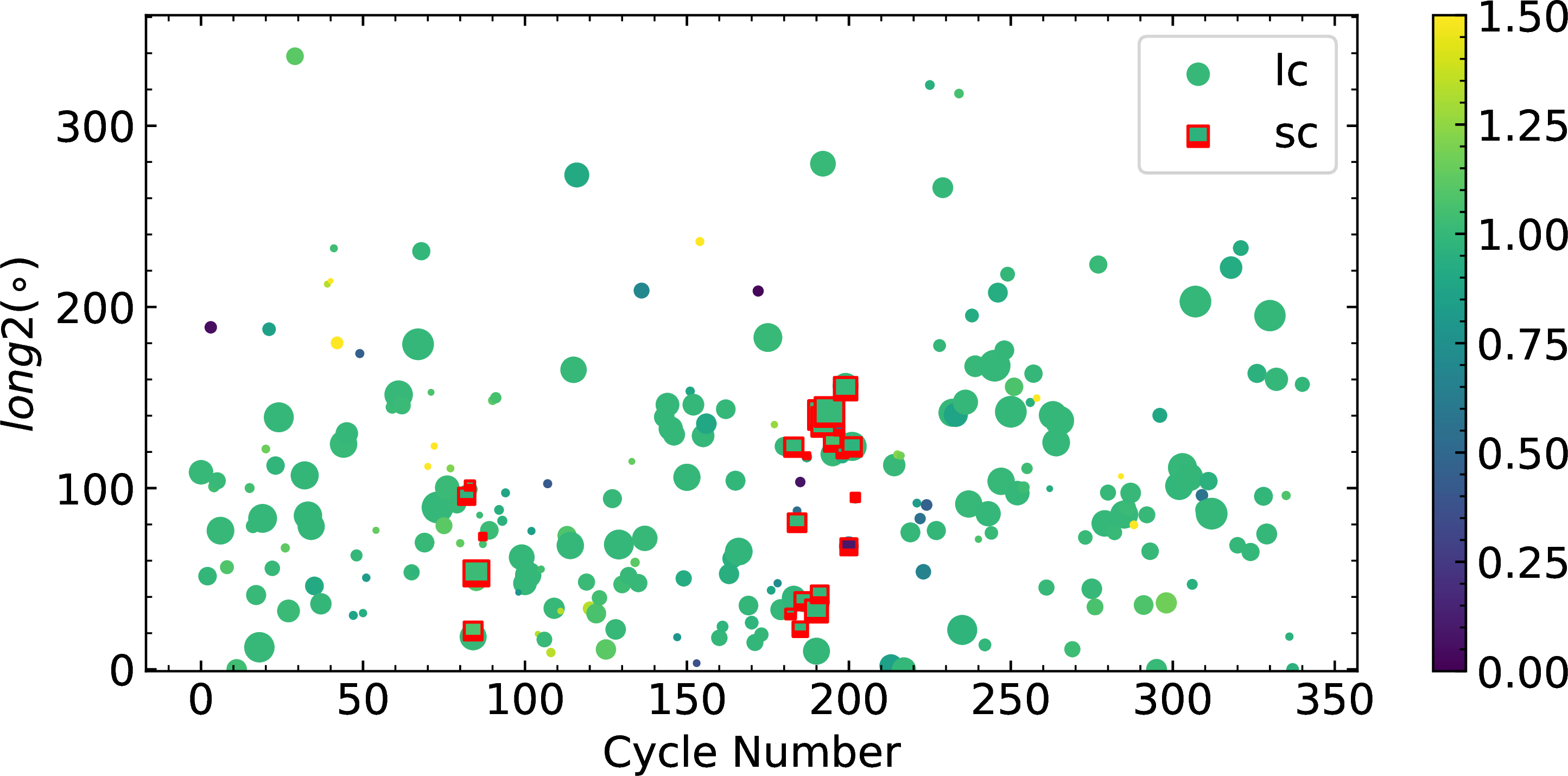}
\caption{Variations of the longitudes of spot/facula versus the cycle number. The upper panel represents the secondary star, and the lower panel is for the primary star. The color bars represent the contrast of spot/facula.}
\label{fig:longitude}
\end{figure}

From Figure \ref{fig:smooth}, one can clearly see the O'Connell effect in this binary system, i.e. the two out-of-eclipse maxima are unequally high. It can also be seen in the residuals of the binary model solution without spot/facula for each cycle of light curve (the upper left panels in Figs. \ref{fig:lcmodel} and \ref{fig:scmodel}). There are several explanations for this effect, the most common of which is the spot modulation \citep{2009SASS...28..107W}. We have tested the single spot model, but it can not well suite to the out-of-eclipse light curve whether on the primary or the secondary star as some trend still exists in the residuals (see the upper right and lower left panels in Figs. \ref{fig:lcmodel} and \ref{fig:scmodel}). We have tried to optimize the other parameters like $q$, $a\mathrm{sin}i$, albedo and reflection coefficients, but it hasn't improved much. Thus we have constructed the model with two spots, one for each component star, for which there is no trend and with the smallest $rms$ scatter of the residuals. The facula can also be called as hot spot, the temperature of which is higher than that of the photosphere around it. We optimize the four parameters of each spot/facula, i.e. the colatitude $colat$, the longitude $long$, the size $radius$, and the contrast $relteff$, along with the binary parameters in Section \ref{sec:phoebe} with the narrower search range for each cycle light curve. The bounds of parameters for optimization are $e\mathrm{cos}\omega \in$ (0.002, 0.008), $e\mathrm{sin}\omega \in$ (-0.03, 0.001), $i \in$ (84.7, 89.7), $r_{teff} \in$ (0.85, 1.0), $R_1 \in$ (1.37, 1.77), $R_2 \in$ (0.88,1.28), $colat \in$ (0, 180), $long \in$ (0, 360), $radius \in$ (0, 90), and $relteff \in$ (0, 1.5). As can be seen in the lower right panels in Figures \ref{fig:lcmodel} and \ref{fig:scmodel}, the optimized model solution with two spots suite well to the light curves. The average values of those parameters and uncertainties are listed in Table \ref{tab:binary_spot} for long and short cadence data, respectively. Since the results based on the long and short cadence data are very consistent to each other, we take the parameters from the long cadence data as the final result, of which the uncertainties are slightly smaller. The orbital eccentricity is $e=0.0217\pm0.0008$. The orbital inclination is $i=87.71\pm 0.04^\circ$. The angle of periastron is $\omega=284.1\pm 0.5^\circ$. The masses and radii of the primary and secondary star are $M_1=1.3467 \pm 0.0001\ M_\odot$, $R_1=1.569 \pm 0.003\ R_\odot$, and $M_2=1.0940 \pm 0.0001\ M_\odot$, $R_2=1.078 \pm 0.002\ R_\odot$, respectively. The temperature ratio is $r_{teff}=0.924 \pm 0.001$, so the temperature of the secondary star can be derived as $T_{eff2}=5936\pm25$ K. Although affected by the degeneracy of spot parameters $colat$, size $radius$ and contrast $relteff$, it hints that the temperature of spot on the primary star is very close to or slightly higher than that of the photosphere, and it is lower than that of the photosphere on the secondary star. So it may be facula (hot spot) dominant for the primary star, and cold spot dominant for the secondary star. As shown in Figure \ref{fig:longitude}, it seems like that there is some signature of the migration of spot longitudes from one hemisphere to the opposite side on the secondary star, from about cycle 30 to 120, cycle 150 to 250 and cycle 260 to 330, respectively, which means that active regions are formed and disappeared in different longitudes. \cite{2020CoSka..50..481B} also found the longitudinal spot migration on eclipsing binary KIC9821078, but the drift direction was toward increasing longitudes. But we can not see this kind of spot migration on the primary star.

\subsection{Comparison with spots of single stars}
\label{subsec:comparison}
We compare the spot/facula activity on KIC 8098300 with that of 1774 single main sequence (MS) stars analyzed by \cite{2017MNRAS.472.1618G} based on the $\it{Kepler}$ light curves. As shown in the left panel of Figure \ref{fig:rav_rms}, the rms of the spot/facula on this binary is located at the bottom edge in the corresponding temperature range, which means the spot/facula activity in this binary is relatively weaker than that of most single MS stars. As shown in the right panel of Figure \ref{fig:rav_rms}, the lifetime of active region $\tau_{AR} $ on this binary is located near the upper edge in the same temperature range, which implies that the lifetime of the spots/faculae on this binary is longer than that of most single MS stars. This is similar to the case of the eclipsing binary KIC 5359678 \citep{2021MNRAS.504.4302W}. \cite{2017ApJ...851..116M} analyzed a sample of 4000 Sun-like stars with measured rotation periods and found a transition between spot-dominated and facula-dominated variability between rotation periods of 15 and 25 days, which implied that the transition between the two modes is complete for MS stars at the age of the Sun. For this short period binary, it may be facula-dominated for the primary star, and spot-dominated for the secondary star. So the binary may be undergoing the transition between the two modes.

\begin{figure}[htb!]
\centering
\includegraphics[width=0.46\textwidth,height=0.26\textheight, angle=0]{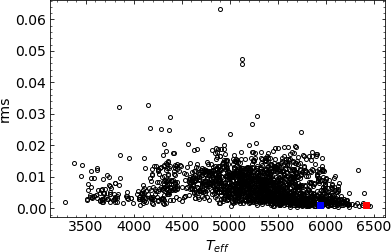}
\includegraphics[width=0.46\textwidth,height=0.26\textheight, angle=0]{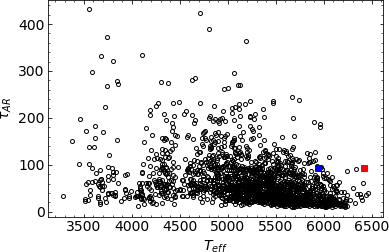}
\caption{Comparison the spot/facula activity on KIC 8098300 with that of spots on the single MS stars in the sample of \cite{2017MNRAS.472.1618G}. The red filled square represent the primary star, while the blue filled square is the secondary star. The black open circles are the sample of single MS stars.}
\label{fig:cmpspots}
\end{figure}

\section{Summary}
\label{sec:summary}

The LAMOST spectra and $\it{Kepler}$ light curves are combined to study the detached eclipsing binary KIC 5359678, which shows the O'Connell effect caused by spot modulation. The tomographic spectra disentangling technology is applied to decompose the time series of spectra and derive the RVs. Thus the parameters of $q = K_1/K_2 = 0.812 \pm 0.007$ and $a\mathrm{\sin}i= 14.984 \pm 0.048\ R_\odot$ are obtained by orbital fitting of RVs. We construct and optimize the binary model with the spot and facula using the PHOEBE code and $\it{curve\_fit}$ function in $\it{Scipy}$ package for each cycle of the light curves. Each parameter for all 264 cycles is averaged as the final value. The orbital eccentricity, the orbital inclination, and the angle of periastron are obtained as $e=0.0217 \pm 0.0008$, $i=87.71 \pm 0.04^\circ$, and $\omega=284.1 \pm 0.5^\circ$, respectively. The masses and radii of the primary and secondary star are $M_1=1.3467 \pm 0.0001\ M_\odot$, $R_1=1.569 \pm 0.003\ R_\odot$, and $M_2=1.0940 \pm 0.0001\ M_\odot$, $R_2=1.078 \pm 0.002\ R_\odot$, respectively. The temperature ratio of the secondary and the primary star is $r_{teff}=0.924 \pm 0.001$, so the temperature of the secondary star can be derived as $T_{eff2}=5936 \pm 25\ $ K. We also find the periastron  precession of $0.000024\pm 0.000001$ d cycle$^{-1}$. It may be spot-dominated for the secondary star and facula-dominated for the primary star. This binary system may be undergoing the transition between the two modes. It shows some signature of spot migration toward decreasing longitudes on the secondary star, but it can not be seen on the primary star.

The residuals of out-of-eclipse are analyzed using ACF and DFT to determine the lifetimes and the rotation periods of active regions. The lifetimes of active regions are longer than that of most single MS stars in the same temperature range. The average rotation period $P_{rot}=4.32$ d is slightly longer than the orbital period (4.3059177 d), which may be caused by either the migration of spots/faculae in longitude or the latitudinal differential rotation. The spot/facula activity in this binary is relatively weaker (with smaller $rms$) than that of most single MS stars.

\begin{acknowledgements}
We would like to thank the referees for the valuable and constructive comments. The authors acknowledge the support from the National Natural Science Foundation of China (NSFC) through the grants 11833002, 12090040, and 12090042. Guoshoujing Telescope (the Large Sky Area Multi-Object Fiber Spectroscopic Telescope LAMOST) is a National Major Scientific Project built by the Chinese Academy of Sciences. Funding for the project has been provided by the National Development and Reform Commission. LAMOST is operated and managed by the National Astronomical Observatories, Chinese Academy of Sciences. This paper includes data collected by the $\it{Kepler}$ mission. Funding for the $\it{Kepler}$  mission  is  provided by the NASA Science Mission directorate. Data presented in this paper were obtained from the Mikulski Archive for Space Telescopes (MAST).
\end{acknowledgements}


\bibliographystyle{raa}
\bibliography{bibtex}

\begin{thebibliography}{63}
\providecommand\natexlab[1]{#1}
\providecommand\JournalTitle[1]{#1}

\bibitem[{Bagnuolo} \& {Gies}(1991)]{1991ApJ...376..266B}
{Bagnuolo}, William~G., J., \& {Gies}, D.~R. 1991, \apj, 376, 266

\bibitem[{Bahar} {et~al.}(2020)]{2020CoSka..50..481B}
{Bahar}, E., {{\"O}zavc{\i}}, {\.I}., \& {{\c{S}}enavc{\i}}, H.~V. 2020,
  Contributions of the Astronomical Observatory Skalnate Pleso, 50, 481

\bibitem[{Balona} \& {Abedigamba}(2016)]{2016MNRAS.461..497B}
{Balona}, L.~A., \& {Abedigamba}, O.~P. 2016, \mnras, 461, 497

\bibitem[{Barnes} {et~al.}(2002)]{2002AN....323..333B}
{Barnes}, J.~R., {James}, D.~J., \& {Collier Cameron}, A. 2002, Astronomische
  Nachrichten, 323, 333

\bibitem[{Basri} {et~al.}(2011)]{2011AJ....141...20B}
{Basri}, G., {Walkowicz}, L.~M., {Batalha}, N., {et~al.} 2011, \aj, 141, 20

\bibitem[{Borucki} {et~al.}(2010)]{2010Sci...327..977B}
{Borucki}, W.~J., {Koch}, D., {Basri}, G., {et~al.} 2010, Science, 327, 977

\bibitem[{Collier Cameron}(1995)]{1995MNRAS.275..534C}
{Collier Cameron}, A. 1995, \mnras, 275, 534

\bibitem[{Conroy} {et~al.}(2020)]{2020ApJS..250...34C}
{Conroy}, K.~E., {Kochoska}, A., {Hey}, D., {et~al.} 2020, \apjs, 250, 34

\bibitem[{Cui} {et~al.}(2012)]{Cui2012}
{Cui}, X.-Q., {Zhao}, Y.-H., {Chu}, Y.-Q., {et~al.} 2012, Research in Astronomy
  and Astrophysics, 12, 1197

\bibitem[{Czesla} {et~al.}(2019)]{2019A&A...623A.107C}
{Czesla}, S., {Terzenbach}, S., {Wichmann}, R., \& {Schmitt}, J.~H.~M.~M. 2019,
  \aap, 623, A107

\bibitem[{De Cat} {et~al.}(2015)]{2015ApJS..220...19D}
{De Cat}, P., {Fu}, J.~N., {Ren}, A.~B., {et~al.} 2015, \apjs, 220, 19

\bibitem[{Foukal} {et~al.}(2006)]{2006Natur.443..161F}
{Foukal}, P., {Fr{\"o}hlich}, C., {Spruit}, H., \& {Wigley}, T.~M.~L. 2006,
  \nat, 443, 161

\bibitem[{Foukal} \& {Vernazza}(1979)]{1979ApJ...234..707F}
{Foukal}, P., \& {Vernazza}, J. 1979, \apj, 234, 707

\bibitem[{Fu} {et~al.}(2020)]{2020RAA....20..167F}
{Fu}, J.-N., {Cat}, P.~D., {Zong}, W., {et~al.} 2020, Research in Astronomy and
  Astrophysics, 20, 167

\bibitem[{Giles} {et~al.}(2017)]{2017MNRAS.472.1618G}
{Giles}, H. A.~C., {Collier Cameron}, A., \& {Haywood}, R.~D. 2017, \mnras,
  472, 1618

\bibitem[{Gu} {et~al.}(2003)]{2003A&A...405..763G}
{Gu}, S.~H., {Tan}, H.~S., {Wang}, X.~B., \& {Shan}, H.~G. 2003, \aap, 405, 763

\bibitem[{Hadrava}(1995)]{1995A&AS..114..393H}
{Hadrava}, P. 1995, \aaps, 114, 393

\bibitem[{Hathaway}(2010)]{2010LRSP....7....1H}
{Hathaway}, D.~H. 2010, Living Reviews in Solar Physics, 7, 1

\bibitem[{Hensberge} {et~al.}(2008)]{2008A&A...482.1031H}
{Hensberge}, H., {Iliji{\'c}}, S., \& {Torres}, K.~B.~V. 2008, \aap, 482, 1031

\bibitem[{Hou} {et~al.}(2018)]{2018SPIE10702E..1IH}
{Hou}, Y., {Tang}, L., {Xu}, M., {et~al.} 2018, in Society of Photo-Optical
  Instrumentation Engineers (SPIE) Conference Series, Vol. 10702, Ground-based
  and Airborne Instrumentation for Astronomy VII, ed. C.~J. {Evans},
  L.~{Simard}, \& H.~{Takami}, 107021I

\bibitem[{Iglesias-Marzoa} {et~al.}(2015)]{2015PASP..127..567I}
{Iglesias-Marzoa}, R., {L{\'o}pez-Morales}, M., \& {Jes{\'u}s Ar{\'e}valo
  Morales}, M. 2015, \pasp, 127, 567

\bibitem[{Ilijic} {et~al.}(2004)]{2004ASPC..318..111I}
{Ilijic}, S., {Hensberge}, H., {Pavlovski}, K., \& {Freyhammer}, L.~M. 2004, in
  Astronomical Society of the Pacific Conference Series, Vol. 318,
  Spectroscopically and Spatially Resolving the Components of the Close Binary
  Stars, ed. R.~W. {Hilditch}, H.~{Hensberge}, \& K.~{Pavlovski}, 111

\bibitem[{Jones} {et~al.}(2020)]{2020ApJS..247...63J}
{Jones}, D., {Conroy}, K.~E., {Horvat}, M., {et~al.} 2020, \apjs, 247, 63

\bibitem[{Kirk} {et~al.}(2016)]{2016AJ....151...68K}
{Kirk}, B., {Conroy}, K., {Pr{\v{s}}a}, A., {et~al.} 2016, \aj, 151, 68

\bibitem[{Liu} {et~al.}(2020)]{2020arXiv200507210L}
{Liu}, C., {Fu}, J., {Shi}, J., {et~al.} 2020, arXiv e-prints, arXiv:2005.07210

\bibitem[{Lomb}(1976)]{1976Ap&SS..39..447L}
{Lomb}, N.~R. 1976, \apss, 39, 447

\bibitem[{Luo} {et~al.}(2004)]{Luo2004}
{Luo}, A.~L., {Zhang}, Y.-X., \& {Zhao}, Y.-H. 2004, Society of Photo-Optical
  Instrumentation Engineers (SPIE) Conference Series, Vol. 5496, {Design and
  implementation of the spectra reduction and analysis software for LAMOST
  Telescope}, ed. H.~{Lewis} \& G.~{Raffi}, Society of Photo-Optical
  Instrumentation Engineers (SPIE) Conference Series, Vol. 5496, \procspie, ed.
  H.~{Lewis} \& G.~{Raffi}, 756

\bibitem[{Luo} {et~al.}(2012)]{2012RAA....12.1243L}
{Luo}, A.~L., {Zhang}, H.-T., {Zhao}, Y.-H., {et~al.} 2012, Research in
  Astronomy and Astrophysics, 12, 1243

\bibitem[{Lurie} {et~al.}(2017)]{2017AJ....154..250L}
{Lurie}, J.~C., {Vyhmeister}, K., {Hawley}, S.~L., {et~al.} 2017, \aj, 154, 250

\bibitem[{Maunder}(1904)]{1904MNRAS..64..747M}
{Maunder}, E.~W. 1904, \mnras, 64, 747

\bibitem[{Maxted}(2016)]{2016A&A...591A.111M}
{Maxted}, P.~F.~L. 2016, \aap, 591, A111

\bibitem[{McQuillan} {et~al.}(2013)]{2013MNRAS.432.1203M}
{McQuillan}, A., {Aigrain}, S., \& {Mazeh}, T. 2013, \mnras, 432, 1203

\bibitem[{McQuillan} {et~al.}(2014)]{2014ApJS..211...24M}
{McQuillan}, A., {Mazeh}, T., \& {Aigrain}, S. 2014, \apjs, 211, 24

\bibitem[{Montet} {et~al.}(2017)]{2017ApJ...851..116M}
{Montet}, B.~T., {Tovar}, G., \& {Foreman-Mackey}, D. 2017, \apj, 851, 116

\bibitem[{Mowlavi} {et~al.}(2014)]{2014bsee.confP..23M}
{Mowlavi}, N., {Holl}, B., {Siopis}, C., \& {Geneva Gaia CU7 Team Members}.
  2014, in Binary Systems, their Evolution and Environments, P2

\bibitem[{Nielsen} {et~al.}(2019)]{2019A&A...622A..85N}
{Nielsen}, M.~B., {Gizon}, L., {Cameron}, R.~H., \& {Miesch}, M. 2019, \aap,
  622, A85

\bibitem[O'Connell(1951)]{Connell51}
O'Connell, D. 1951, Publications of the Riverview College Observatory, 2, 85

\bibitem[{{\"O}zavc{\i}} {et~al.}(2018)]{2018MNRAS.474.5534O}
{{\"O}zavc{\i}}, I., {{\c{S}}enavc{\i}}, H.~V., {I{\c{s}}{\i}k}, E., {et~al.}
  2018, \mnras, 474, 5534

\bibitem[{Pan} {et~al.}(2020)]{2020ApJ...905...67P}
{Pan}, Y., {Fu}, J.-N., {Zong}, W., {et~al.} 2020, \apj, 905, 67

\bibitem[{Perryman} {et~al.}(2001)]{2001A&A...369..339P}
{Perryman}, M.~A.~C., {de Boer}, K.~S., {Gilmore}, G., {et~al.} 2001, \aap,
  369, 339

\bibitem[{Pi} {et~al.}(2019)]{2019ApJ...877...75P}
{Pi}, Q.-f., {Zhang}, L.-y., {Bi}, S.-l., {et~al.} 2019, \apj, 877, 75

\bibitem[{Press} \& {Rybicki}(1989)]{1989ApJ...338..277P}
{Press}, W.~H., \& {Rybicki}, G.~B. 1989, \apj, 338, 277

\bibitem[{Pr{\v{s}}a} \& {Zwitter}(2005)]{2005ApJ...628..426P}
{Pr{\v{s}}a}, A., \& {Zwitter}, T. 2005, \apj, 628, 426

\bibitem[{Pr{\v{s}}a} {et~al.}(2011)]{2011AJ....141...83P}
{Pr{\v{s}}a}, A., {Batalha}, N., {Slawson}, R.~W., {et~al.} 2011, \aj, 141, 83

\bibitem[{Pr{\v{s}}a} {et~al.}(2016)]{2016ApJS..227...29P}
{Pr{\v{s}}a}, A., {Conroy}, K.~E., {Horvat}, M., {et~al.} 2016, \apjs, 227, 29

\bibitem[{Reinhold} {et~al.}(2017)]{2017A&A...603A..52R}
{Reinhold}, T., {Cameron}, R.~H., \& {Gizon}, L. 2017, \aap, 603, A52

\bibitem[{Ricker} {et~al.}(2015)]{2015JATIS...1a4003R}
{Ricker}, G.~R., {Winn}, J.~N., {Vanderspek}, R., {et~al.} 2015, Journal of
  Astronomical Telescopes, Instruments, and Systems, 1, 014003

\bibitem[{Scargle}(1982)]{1982ApJ...263..835S}
{Scargle}, J.~D. 1982, \apj, 263, 835

\bibitem[{Simon} \& {Sturm}(1994)]{1994A&A...281..286S}
{Simon}, K.~P., \& {Sturm}, E. 1994, \aap, 281, 286

\bibitem[{Slawson} {et~al.}(2011)]{2011AJ....142..160S}
{Slawson}, R.~W., {Pr{\v{s}}a}, A., {Welsh}, W.~F., {et~al.} 2011, \aj, 142,
  160

\bibitem[{Sterken}(2005)]{2005ASPC..335....3S}
{Sterken}, C. 2005, in Astronomical Society of the Pacific Conference Series,
  Vol. 335, The Light-Time Effect in Astrophysics: Causes and cures of the O-C
  diagram, ed. C.~{Sterken}, 3

\bibitem[{Strassmeier}(2009)]{2009A&ARv..17..251S}
{Strassmeier}, K.~G. 2009, \aapr, 17, 251

\bibitem[{Stumpe} {et~al.}(2012)]{2012PASP..124..985S}
{Stumpe}, M.~C., {Smith}, J.~C., {Van Cleve}, J.~E., {et~al.} 2012, \pasp, 124,
  985

\bibitem[{VanderPlas}(2018)]{2018ApJS..236...16V}
{VanderPlas}, J.~T. 2018, \apjs, 236, 16

\bibitem[{Vogt} \& {Penrod}(1983)]{1983PASP...95..565V}
{Vogt}, S.~S., \& {Penrod}, G.~D. 1983, \pasp, 95, 565

\bibitem[{Wang} {et~al.}(2021)]{2021MNRAS.504.4302W}
{Wang}, J., {Fu}, J., {Niu}, H., {et~al.} 2021, \mnras, 504, 4302

\bibitem[{Wilsey} \& {Beaky}(2009)]{2009SASS...28..107W}
{Wilsey}, N.~J., \& {Beaky}, M.~M. 2009, Society for Astronomical Sciences
  Annual Symposium, 28, 107

\bibitem[Wu {et~al.}(2014)]{Wu2014}
Wu, Y., Du, B., Luo, A., Zhao, Y., \& Yuan, H. 2014, Proceedings of the
  International Astronomical Union, 10, 340

\bibitem[{Xiang} {et~al.}(2020)]{2020MNRAS.492.3647X}
{Xiang}, Y., {Gu}, S., {Wolter}, U., {et~al.} 2020, \mnras, 492, 3647

\bibitem[{Zechmeister} \& {K{\"u}rster}(2009)]{2009A&A...496..577Z}
{Zechmeister}, M., \& {K{\"u}rster}, M. 2009, \aap, 496, 577

\bibitem[{Zhao} {et~al.}(2012)]{2012RAA....12..723Z}
{Zhao}, G., {Zhao}, Y.-H., {Chu}, Y.-Q., {Jing}, Y.-P., \& {Deng}, L.-C. 2012,
  Research in Astronomy and Astrophysics, 12, 723

\bibitem[{Zong} {et~al.}(2018)]{2018ApJS..238...30Z}
{Zong}, W., {Fu}, J.-N., {De Cat}, P., {et~al.} 2018, \apjs, 238, 30

\bibitem[{Zong} {et~al.}(2020)]{2020ApJS..251...15Z}
{Zong}, W., {Fu}, J.-N., {De Cat}, P., {et~al.} 2020, \apjs, 251, 15

\end{thebibliography}

\label{lastpage}

\end{document}